\documentclass[journal]{IEEEtran}

\usepackage[figuresright]{rotating}
\usepackage{comment}
\usepackage[hang,tight]{subfigure}
\usepackage{amsmath}
\usepackage{bm}
\usepackage{booktabs}
\usepackage{amsfonts}

\usepackage{multirow}
\usepackage{algorithm}
\usepackage{algorithmic}

\usepackage{tabularx}
\usepackage{xcolor}

\usepackage{enumitem}

\begin{document}

%\title{Contrastive Learning-based Activity Recognition using WiFi Channel State Information}
%\title{Contrastive Self-Supervised Learning-based Cross-Subject Activity Recognition using WiFi CSI}
%\title{Contrastive Self-Supervised Learning-based Activity Recognition using WiFi CSI}

%\title{Contrastive Self-Supervised Learning-based Human Activity Recognition With diffusion models}

%\title{Diffusion Model-based Contrastive Self-Supervised Learning for   Activity Recognition using WiFi CSI}

%\title{Diffusion Model-based Contrastive Self-Supervised Learning for Human Activity Recognition}
\title{Diffusion Model-based Contrastive Learning for Human Activity Recognition}

%chunjingxiao@gmail.com.
%Yane Hou, houyane@henu.edu.cn
\author{
    Chunjing Xiao,
    Yanhui Han,
    Wei Yang,
    Yane Hou,
    Fangzhan Shi,
    Kevin Chetty
\thanks{Manuscript received 4 June 2024; accepted 10 July 2024. This work is supported in part by the National Natural Science Foundation of China under Grant 41801310.
\emph{(Corresponding author: Wei Yang.)} }
    
\thanks{ Chunjing Xiao is with the School of Computer and Information Engineering, Henan University, Kaifeng 475004, China, and also with University College London, London WC1E 6BT, UK (e-mail: ChunjingXiao@gmail.com).}
\thanks{ Yanhui Han, Wei Yang and Yane Hou are with the School of Computer and Information Engineering, Henan University, and also with the Henan Key Laboratory of Big Data Analysis and Processing, Henan University, Kaifeng 475004, China (e-mail: hanyanhui@henu.edu.cn, yangwei@henu.edu.cn, houyane@henu.edu.cn).}
\thanks{ Fangzhan Shi and Kevin Chetty are with the Department of Security and Crime Science, University College London, London WC1E 6BT, UK (e-mail: Fangzhan.shi.17@ucl.ac.uk, k.chetty@ucl.ac.uk).} 
}

%\protect\\ (Corresponding author: xxxx.)

%can transfer knowledge from the source domain to the target domain

% The paper headers
\markboth{Journal of \LaTeX\ Class Files,~Vol.~14, No.~8, August~2015}%
{Shell \MakeLowercase{\textit{et al.}}: Bare Demo of IEEEtran.cls for IEEE Journals}

\newcommand{\tcr}{\textcolor{red}}
\newcommand{\tcb}{\textcolor{blue}}

\maketitle

\begin{abstract}
WiFi Channel State Information (CSI)-based activity recognition has sparked numerous studies due to its widespread availability and privacy protection. However, when applied in practical applications, general CSI-based recognition models may face challenges related to the limited generalization capability, since individuals with different behavior habits will cause various fluctuations in CSI data and it is difficult to gather enough training data to cover all kinds of motion habits. To tackle this problem, we design a diffusion model-based Contrastive Learning framework for human Activity Recognition (CLAR) using WiFi CSI. On the basis of the contrastive learning framework, we primarily introduce two components for CLAR to enhance CSI-based activity recognition. To generate diverse augmented data and complement limited training data, we propose a diffusion model-based time series-specific augmentation model. In contrast to typical diffusion models that directly apply conditions to the generative process, potentially resulting in distorted CSI data, our tailored model dissects these condition into the high-frequency and low-frequency components, and then applies these conditions to the generative process with varying weights. This can alleviate data distortion and yield high-quality augmented data. To efficiently capture the difference of the sample importance, we present an adaptive weight algorithm. Different from typical contrastive learning methods which equally consider all the training samples, this algorithm adaptively adjusts the weights of positive sample pairs for learning better data representations. The experiments suggest that CLAR achieves significant gains compared to state-of-the-art methods.
\end{abstract}

\begin{IEEEkeywords}
Contrastive learning, self-supervised learning, diffusion probabilistic models, WiFi CSI, activity recognition.
\end{IEEEkeywords}

\section{Introduction}
\label{sec:intro1}

%These models are typically powered by supervised machine learning algorithms. In order for an adaptive model to maintain an acceptable performance, a large training dataset is needed, which makes the training phase time consuming, labor intensive, and expensive. Data collection has been identified as a major obstacle in personalized and precision medicine [5].
%----Personalized Activity Recognition using Partially Available Target Data, TMC 2022

%Several WiFi sensing techniques have the potential for achieving non-restrictive, privacy-friendly indoor activity sensing when compared to current technologies such as cameras or wearable sensors
%----Self-Supervised WiFi-Based Activity Recognition, arxiv 2021

Human activity recognition is considered a key aspect for a variety of real-world applications, 
%------------------------------rpt-------------------
%Human activity recognition is considered a key aspect for a variety of applications,
such as health monitoring and smart home~\cite{gu2021survey,yang2022review}.
Among a great many recognition techniques, WiFi Channel State Information (CSI)-based approaches have the potential to achieve device-free, non-intrusive and privacy-friendly activity sensing, when compared to camera-based or wearable sensor-based methods~\cite{yousefi2017survey,ma2019wifi,zhang2020device}.
%----Exploring Contrastive Learning in Human Activity Recognition for Healthcare, nips 2020
However, when applying CSI-based activity recognition methods to practical applications, one  major hurdle is the difficulty of gathering labeled samples~\cite{gu2021survey,xiao2022self}. 
%Since CSI data abstract and difficult to interpret by humans, labeled data collection usually require protocol-based tasks in laboratory settings.
%------------------------------rpt-------------------
Due to the abstract and challenging nature of CSI data interpretation for humans, collecting labeled data often requires protocol-based tasks in controlled laboratory environments.

Contrastive learning can be a potential solution to overcome the limitations associated with the lack of labels, because it can effectively leverage an enormous number of unlabelled samples to train the model without using labels~\cite{2022Assessing}.
It has shown superior performance in the image~\cite{2016Pixel,SimCLR2020}, natural language and time-series data processing~\cite{2022GL-CLeF,deldari2022cocoa}. However, directly applying contrastive learning to CSI-based activity recognition tasks is confronted with additional issues.

\begin{figure}[tb]
	\centering
	\subfigure[Gaussian blur]{\label{fig:blur}\includegraphics[width=0.241\textwidth]{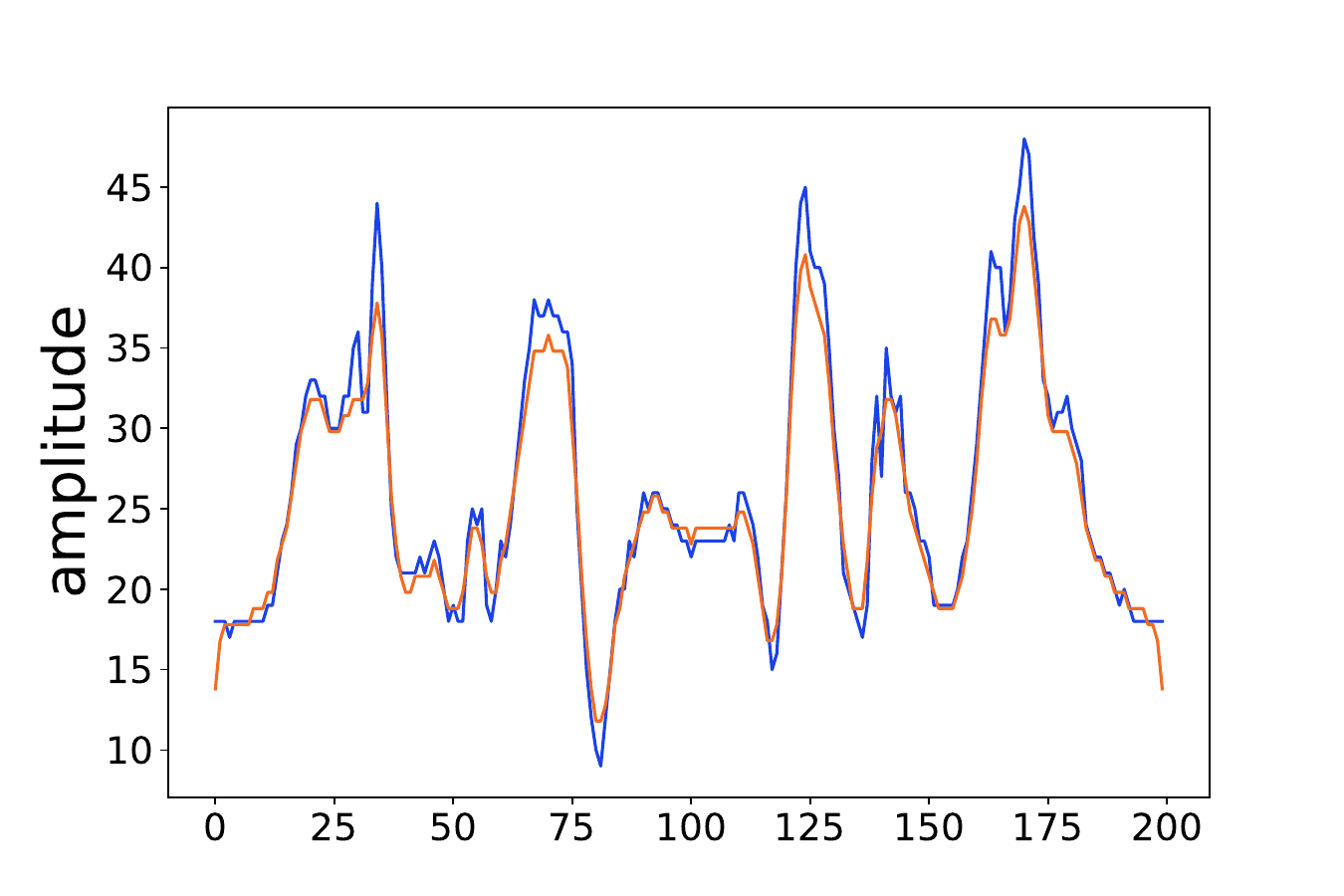}}
	\subfigure[DDPM-based augmentation]{\label{fig:inject}\includegraphics[width=0.241\textwidth]{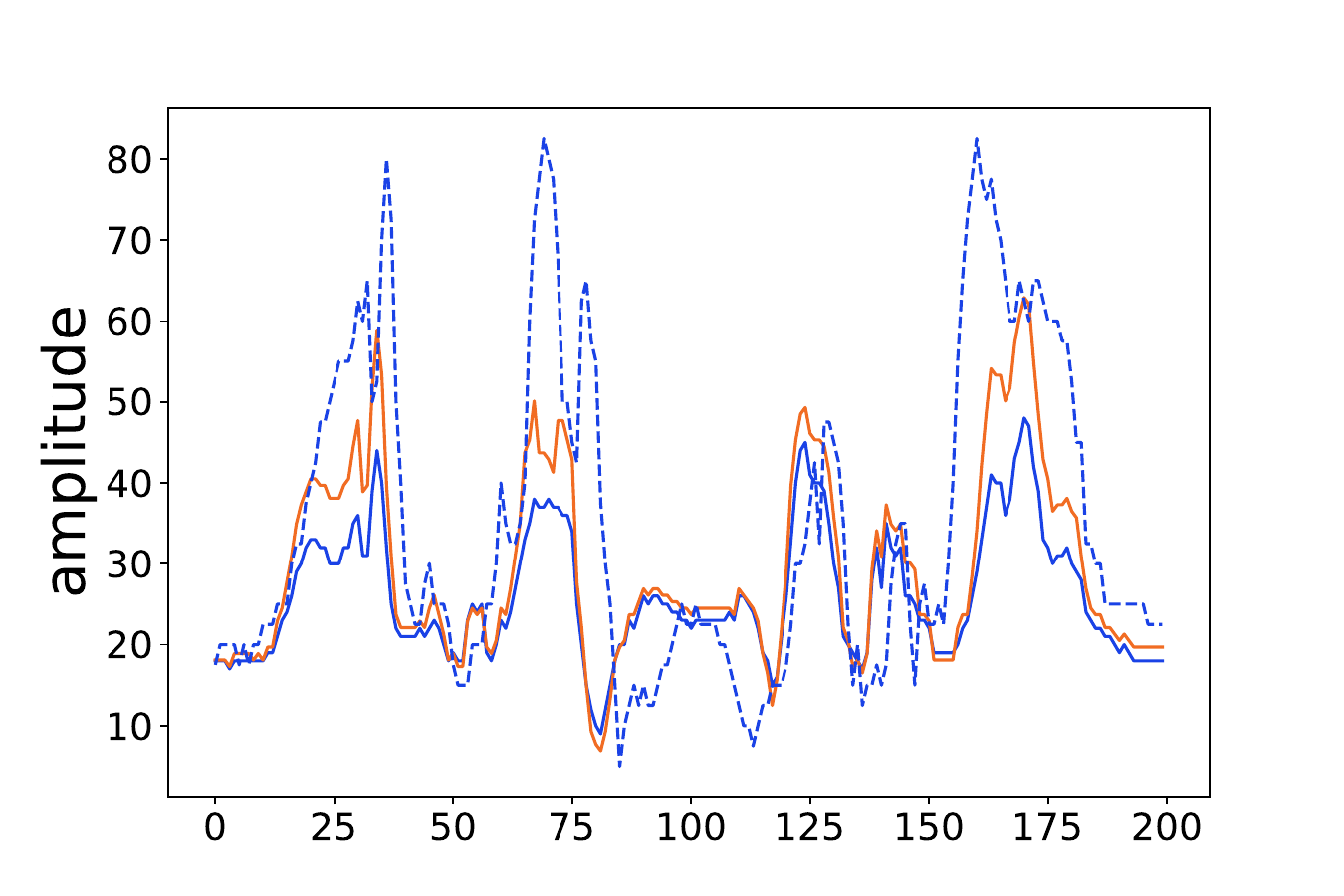}}
	%\vspace{-0.3cm}
	\caption{Augmented data by different methods. The orange line denotes the augmented data and the blue line refers to the real one. (a) The  waveform of augmented data by Gaussian blur (orange) is almost the same to the original one (blue). (b) The waveform of augmented data  by our DDPM-based augmentation method (orange) can combine the characteristics of the two samples (solid and dotted blue).}
	\label{fig:Pretexttasks}
	\vspace{-0.3cm}
\end{figure}

%----What Should Not Be Contrastive in Contrastive Learning, ICLR 2021
%Contrastive learning methods have been able to produce impressive transferable visual representations by learning to be invariant to different data augmentations
%----Dual-Stream Contrastive Learning for Channel State Information Based Human Activity Recognition, IEEE Journal of Biomedical and Health Informatics 2023.

First, prevailing data augmentation approaches in contrastive learning cannot effectively generate diverse augmented samples for CSI data.
General augmentation methods (e.g., Gaussian blur and color distortion), which are particularly designed for image and text data, hardly change the waveform shape of CSI (a kind of time-series data) due to differences in data structure. 
%------------------------------rpt-------------------
%General augmentation methods (e.g., Gaussian blur and color distortion), which are primarily tailored for image and text data, have limited impact on the waveform shape of CSI (a type of time-series data) due to structural differences. 
%which focus on manipulating pixels to generate augmented data.
An example is presented in Figure~\ref{fig:blur}, which suggests that the augmented waveform by Gaussian blur (orange) is quite similar to the original one (blue).
However, highly similar augmented samples can provide few benefits for performance improvement~\cite{2021Improving}.  
On the other hand, CSI-based activity recognition requires diverse augmented data to enhance generalization capacity.
Since waveforms of CSI data collected from users with different motion habits can be different even if they perform the same action, it is difficult to gather enough training data to cover all kinds of motion habits ~\cite{2019CsiGAN}.
%Hence, the test data can be out of the distribution of training data, and correspondingly the model trained based on training data might not perform well on test data.
Therefore, augmentation approaches for CSI data should be able to generate augmented data with new motion habits to complement limited training data.

\begin{figure}[tb]
	\centering
	\includegraphics[width=0.40\textwidth]{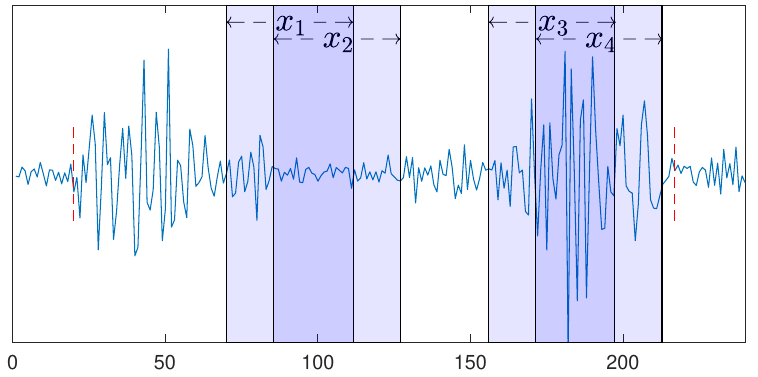}
	\vspace{-0.3cm}
	\caption{Positive sample pairs extracted from an activity where the dotted red lines are the start and end points and there is pause near the center. Compared to positive pair ($x_3$, $x_4$), positive pair ($x_1$, $x_2$) should provide fewer clues for learning data representation because they contain more pause data.}
	\label{fig:pauseExample}
	\vspace{-0.3cm}
\end{figure}

Second, typical contrastive learning models fail to consider the difference of the sample importance during model training. In contrastive learning, the same weights are generally assigned to all the positive sample pairs. However, for CSI-based activity recognition, different positive sample pairs might provide various clues for learning data representation. 
%------------------------------rpt-------------------
%Second, conventional contrastive learning approaches do not account for variations in sample importance during model training. In contrastive learning, equal weights are usually applied to all positive pairs. Nevertheless, for CSI-based activity recognition, different positive pairs could offer diverse insights for enhancing  representations.
%----
For some activities consisting of multiple strokes, such as drawing X and lying down, there might be a pause between two strokes.
If positive sample pairs extracted from CSI data contain more pause data, they will provide fewer clues for learning data representation, and should play a minor role in model training, and vice versa.
An example of drawing X is presented in Figure~\ref{fig:pauseExample}, where the dotted red lines are the real start and end points of the activity. In this activity, there is a pause between the two strokes. Compared to activity pair ($x_3$, $x_4$),  pause pair ($x_1$, $x_2$) contains more pause data and should provide minor clues for learning data representation.
%------------------------------rpt-------------------
%An example of drawing X is presented in Figure~\ref{fig:pauseExample}, where the dotted red lines represent the actual starting and ending points of the action. In the activity, there is a pause between the two strokes.
Compared to activity pair ($x_3$, $x_4$),  pause pair ($x_1$, $x_2$) contains more pause data and should provide minor insights for boosting representations.

%For some activities which consist of multiple strokes, there might be a pause between two strokes, i.e., there exists a piece of data in the absence of activities in an activity sequence. An example of drawing X is presented in Figure~\ref{fig:pauseExample}, where the dotted red lines are the real start and end points of the activity. In the middle of this activity data, there is a pause between the two strokes.
%For contrastive learning, the widely used cropping operation will extract two pieces of data from the same sample to form a positive pair. If the extracted two pieces contain activities data, they can provide more clues for learning data representation, and should play a more important role for model training, and vice versa. For example, in Figure~\ref{fig:pauseExample},  the positive pair, ($x_3$, $x_4$), contains more activity data and should provides more clues, compared to ($x_1$, $x_2$). 

To address these issues, we propose a diffusion model-based Contrastive Learning framework for human Activity Recognition (CLAR) using WiFi CSI.
%------------------------------rpt-------------------
%To tackle the problems, we devise a diffusion model-based Contrastive Learning framework for human Activity Recognition (CLAR) using WiFi CSI.
In this framework, we design two components for the CSI-based activity recognition: a denoising diffusion probabilistic model (DDPM)-based time series-specific data augmentation model and an adaptive weight algorithm. 
The former aims to generate  augmented data with new motion habits. The latter attempts to compute proper weights for different sample pairs. 
The augmented samples and adaptive weights will be used to build the contrastive loss to learn better data representation.

%Different from typical DDPM-based models which aim to generate similar data with given samples~\cite{2020DDPM,song2021denoising}, we try to manufacture new augmented data with different characteristics. 

%The DDPM-based augmentation model takes  a source sample and a reference sample from users with different habits as inputs, and produces a new sample with the combined characteristics of them. 
The DDPM-based augmentation model feeds the prior derived from a source sample into the diffusion model, and regards a reference sample as the condition which is imposed on the reverse diffusion (generative) process of DDPM to generate augmented data with combined characteristics.
%~\cite{2020DDPM,song2021denoising}. In this way, the model can generate a new sample with the combined characteristics.
However, typical imposition ways (i.e., directly imposing the reference sample on the generative process) in the DDPM-based models~\cite{2020DDPM,song2021denoising,rombach2022high,2022BDDM} may cause CSI data distortion.
%However, directly imposing the reference sample into the reverse process might cause data distortion. 
At the earlier steps of the generative process, the generated latent variable is noised and relatively flat. However, the reference data generally has a large fluctuation due to containing activity information~\cite{wang2015understanding}. 
%Directly imposing the reference data might make the data distorted. 
%Hence, directly involving the reference data might make the data disharmonious. 
While, involving the reference data with a big fluctuation on the flat variable might make the data disharmonious.
To address this issue, we first decompose the condition (the reference data) into the two parts: the high-frequency one and the low-frequency one. Then, we adjust the weights of these two conditions according to the reverse diffusion steps, i.e., exert the high-frequency one with the increasing weight and the low-frequency one with the decreasing weight on the generative process.
%larger weights for the low-frequency one and smaller weights for the high-frequency one in earlier iterations.
The reason is that at the earlier steps, the flat latent variables can only involve less high-frequency data with a big fluctuation to avoid data distortion, but can involve more low-frequency data with a small fluctuation. 
This exertion procedure can effectively alleviate the distortion problem and generate effective augmented data to complement the limited training data.
A visual example of the generated sample is presented in Figure~\ref{fig:inject}, where the source sample is from the user with the habit of tending to draw a small circle (solid blue), and the reference sample for a large circle (dotted blue). Correspondingly the generated sample is the one for a middle circle (orange).

The adaptive weight algorithm dynamically computes the weights of positive sample pairs, which are involved in the contrastive loss to boost model performance.
%------------------------------rpt-------------------
%The adaptive weight algorithm dynamically computes the weights of positive  pairs, which are involved in the contrastive loss to boost model performance.
For CSI data, different positive sample pairs provide various clues for learning data representation, i.e., positive sample pairs with fewer activity data should play a minor role in model training since they contain fewer clues for learning data representation and vice versa.
%------------------------------rpt-------------------
%For CSI data, different positive sample pairs contain diverse clues for learning representations, i.e., positive sample pairs with fewer action data should play a minor role in model training since they contain fewer clues for learning data representation and vice versa.
Hence, for each positive sample pair, we first compute a response map to reflect the amount of activity data in the positive pair and then calculate the weight based on the response map. 
This weight will be incorporated into the contrastive loss to enhance the model performance. 
%------------------------------rpt-------------------
%This weight will be incorporated into the contrastive loss to advocate  model performance. 
By incorporating the DDPM-based augmentation model and the adaptive weight algorithm into the basic contrastive learning framework, 
our model can efficiently boost recognition performance.
The key contributions are summarized as follows:
\begin{itemize}
	\item We propose a diffusion model-based Contrastive Learning framework for Activity Recognition, CLAR, which involves augmented data with new characteristics and adaptive sample weights to enhance recognition performance.
	\item We design a DDPM-based time series-specific augmentation model, which decomposes the condition into the high-frequency and low-frequency parts and adopts weight-varying conditions to guide the generative process to generate high-quality augmented data.
    %decomposes the conditions into high-frequency and low-frequency parts and adopt varying weighs to alleviate data distortion for CSI data.
	\item We present an adaptive weight algorithm, which can capture various clues provided in different CSI data to assign appropriate weights for training instances. 
	\item Experiment results based on the two public datasets illustrate that our framework outperforms the state-of-the-art approaches. 
	%------------------------------rpt-------------------
    %\item The experimental results using two public datasets demonstrate that our method surpasses the current state-of-the-art methods.
\end{itemize}

\section{Preliminaries}
\label{sec:Pre}

In this section, we give the necessary background information on the contrastive learning framework and the denoising diffusion probabilistic model.
%------------------------------rpt-------------------
%Here, we give the necessary background  on the contrastive learning framework and the denoising diffusion probabilistic model.

\subsection{Contrastive Learning}
\label{subsec:contrastiveFrame}

Contrastive learning is a type of self-supervised learning algorithm where the goal is to learn feature representations from the inputs by maximizing agreement between differently augmented views of the same data sample via a contrastive loss. When applying contrastive learning to classification tasks, there are two steps during the training process. 
%------------------------------rpt-------------------
%Contrastive learning is a self-supervised learning algorithm that aims to acquire feature representations by maximizing the agreement between differently augmented views of the same data sample through a contrastive loss. When utilizing contrastive learning for classification purposes, the training process typically involves two steps.
%----
The first step is to train a contrastive learning model using unlabeled data. This model will be used to extract high-quality representations from input samples. 
The second step is to train a classifier using a few labeled samples. This classifier will map the representations extracted by the contrastive learning model into corresponding categories.

%----Semi-Supervised Contrastive Learning for Human Activity Recognition, DCOSS 2021
The commonly used contrastive learning model~\cite{SimCLR2020} roughly consists of three components.
($i$) The first part is a data augmentation module, where different augmented views of the same sample are generated with the data augmentation operations. This might involve techniques such as random cropping, color jittering, or random flipping. 
%------------------------------rpt-------------------
%($i$) The data augmentation module seeks to generate different augmented views of the same sample. This might involve techniques such as random cropping, color jittering, or random flipping. 
($ii$) The second part is a feature representation module, which is composed of an encoder and a projection head. This module is responsible for extracting representation vectors from the augmented data examples. Here, the encoder maps the examples into a $d$-dimensional space $\mathbb{R}^d$, and then the projection head $h$ maps $d$-dimensional space into a hyper-spherical (normalized) embedding space as final feature representations. 
%------------------------------rpt-------------------
%($ii$) The feature representation module is composed of an encoder and a projection head. This module is responsible for extracting representations from the augmented examples. Here, the encoder maps the examples into a $d$-dimensional space $\mathbb{R}^d$, and then the projection head $h$ maps $d$-dimensional space into a hyper-spherical (normalized) embedding space as final feature representations. 
($iii$) The last part is the contrastive loss function, which aims at maximizing the agreement between positive pairs of examples. In particular, given a mini-batch of $M$ data examples, the data augmentation would generate $2M$ data points. The two data points derived from the same example are called a pair of positive examples. Given a positive pair, the other $2(M-1)$ augmented examples within a mini-batch are treated as negative examples. Hence, the loss function for a positive pair $(i,j)$ is defined as:
%------------------------------rpt-------------------
%($iii$) The contrastive loss function attempts to maximize the agreement between positive example pairs. When working with a mini-batch of $M$ data examples, data augmentation produces $2M$ data points. A positive pair consists of two data points generated from the same example. Within a mini-batch, the remaining $2(M-1)$ augmented examples are considered as negative examples. Consequently, the loss function for a positive pair $(i,j)$ is defined as follows:
\begin{equation} \begin{aligned}
{\mathcal{L}_{i,j}} =  - \log \frac{{\exp \left( {{{sim\left( {{\textbf{z}_i},{\textbf{z}_j}} \right)} \mathord{\left/
 {\vphantom {{sim\left( {{\textbf{z}_i},{\textbf{z}_j}} \right)} \tau }} \right.
 \kern-\nulldelimiterspace} \tau }} \right)}}{{\sum\nolimits_{k = 1}^{2M} {{{\rm I}_{\left[ {k \ne i} \right]}}\exp \left( {{{sim\left( {{\textbf{z}_i},{\textbf{z}_k}} \right)} \mathord{\left/
 {\vphantom {{sim\left( {{\textbf{z}_i},{\textbf{z}_k}} \right)} \tau }} \right.
 \kern-\nulldelimiterspace} \tau }} \right)} }},
\label{equ:InfoNCE}
\end{aligned} \end{equation}
where $sim(,)$ denotes cosine similarity, $\textbf{z}$ refers to the extracted feature representation, and $\tau$ is a temperature hyper-parameter scaling the distribution of distances.
%------------------------------rpt-------------------
%where $sim(,)$ denotes cosine similarity, $\textbf{z}$ refers to the extracted representation, and $\tau$ is a temperature hyperparameter scaling the distribution of distances.

\subsection{Denoising Diffusion Probabilistic Model}
\label{subsec:ddpm}

Denoising diffusion probabilistic models (DDPM)~\cite{2020DDPM} is a class of generative models that show superior performance in unconditional image generation.
%------------------------------rpt-------------------
%Denoising diffusion probabilistic models (DDPM)~\cite{2020DDPM} belong to a group of generative models which demonstrate exceptional performance in generating images unconditionally.
It learns a Markov Chain which gradually converts a simple distribution (e.g., isotropic Gaussian) into a data distribution. The generative process learns the reverse of the DDPM forward (diffusion) process: a fixed Markov Chain that gradually adds noise to data. Here, each step in the forward process is a Gaussian translation:
%------------------------------rpt-------------------
%This model trains a Markov Chain that gradually transforms a simple distribution, such as an isotropic Gaussian, into a data distribution. The generative process is the reverse of the DDPM forward (diffusion) process, where a Markov Chain systematically introduces noise into the data. In this case, each step in the forward process involves a Gaussian translation:
\begin{equation} \begin{aligned}
q\left( {{\textbf{z}^t}|{\textbf{z}^{t - 1}}} \right): = N\left( {{\textbf{z}^t};\sqrt {1 - {\beta _t}} {\textbf{z}^{t - 1}},{\beta _t}{\rm I}} \right),
\label{equ:forwardprocess}
\end{aligned} \end{equation}
where ${\beta _1}, \ldots ,{\beta _T}$ is a fixed variance schedule (which undergoes a linear increase between 0 and 1) rather than learned
parameters~\cite{2020DDPM}.
%Eq.(1)
Equation~\ref{equ:forwardprocess} is a process finding ${\textbf{z}^t}$ by adding a small Gaussian noise to the latent variable ${\textbf{z}^{t - 1}}$.
According to the work~\cite{2020DDPM}, 
given clean data ${\textbf{z}^0}$, sampling of ${\textbf{z}^t}$ can be expressed in a closed form:
\begin{equation} \begin{aligned}
q\left( {{\textbf{z}^t}|{\textbf{z}^0}} \right): = N\left( {{\textbf{z}^t};\sqrt {{{\bar \alpha }_t}} {\textbf{z}^0},\left( {1 - {{\bar \alpha }_t}} \right){\rm I}} \right),
\end{aligned} \end{equation}
where ${{\bar \alpha }_t}: = \prod\nolimits_{s = 1}^t {{\alpha _s}}$ and ${\alpha _t}: = 1 - {\beta _t}$.
Therefore, ${\textbf{z}^t}$ is expressed as a linear combination of ${\textbf{z}^0}$ and $\varepsilon$:
\begin{equation} \begin{aligned}
{\textbf{z}^t} = \sqrt {{{\bar \alpha }_t}} {\textbf{z}^0} + \sqrt {1 - {{\bar \alpha }_t}} \varepsilon,
\label{equ:forward}
\end{aligned} \end{equation}
where $\varepsilon  \sim N\left( {0,{\rm I}} \right)$ has the same dimensionality as data
${\textbf{z}^0}$ and latent variables ${\textbf{z}^1}, \ldots ,{\textbf{z}^T}$.

Since the reverse of the forward process, $q\left( {{\textbf{z}^{t - 1}}|{\textbf{z}^t}} \right)$, is intractable, DDPM learns parameterized Gaussian transitions ${p_\theta }\left( {{\textbf{z}^{t - 1}}|{\textbf{z}^t}} \right)$.
The generative (or reverse) process has the same functional form~\cite{2015DUL} as the forward process, and it is expressed as a Gaussian transition with learned mean and fixed variance~\cite{2020DDPM}:
\begin{equation} \begin{aligned}
{p_\theta }\left( {{\textbf{z}^{t - 1}}|{\textbf{z}^t}} \right) = N\left( {{\textbf{z}^{t - 1}};{\mu _\theta }\left( {{\textbf{z}^t},t} \right),\sigma _t^2{\rm I}} \right),
\label{equ:reverse}
\end{aligned} \end{equation}
where $\mu _\theta(\cdot)$ refers to the estimated mean of the distribution.
Further, by decomposing ${{\mu _\theta }}$ into a linear combination of ${{\textbf{z}^t}}$ and the noise approximator ${\varepsilon _\theta }$, the generative process is expressed as:
\begin{equation} \begin{aligned}
{\textbf{z}^{t - 1}} = \frac{1}{{\sqrt {{\alpha _t}} }}\left( {{\textbf{z}^t} - \frac{{1 - {\alpha _t}}}{{\sqrt {1 - {{\bar \alpha }_t}} }}{\varepsilon _\theta }\left( {{\textbf{z}^t},t} \right)} \right) + \sigma _t^2\varepsilon,
\label{equ:generative}
\end{aligned} \end{equation}
which $\varepsilon$ is a noise suggesting that each generation step is stochastic.
Here ${{\varepsilon _\theta }}$ represents a neural network with the same input and output dimensions and the noise predicted by the neural network ${{\varepsilon _\theta }}$ in each step is used for the denoising process in Equation~\ref{equ:generative}.
%------------------------------rpt-------------------
%Here ${{\varepsilon _\theta }}$ refers to a neural network with identical input and output dimensions. The noise forecasted by the neural network ${{\varepsilon _\theta }}$ in each step is utilized for the denoising process in Euqation~\ref{equ:generative}.

\section{CLAR Framework}
\label{sec:method1}

\begin{figure}[tb]
	\centering
	\includegraphics[width=0.49\textwidth]{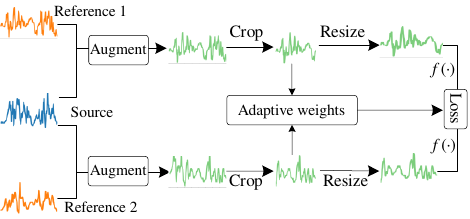}
	\vspace{-0.3cm}
	\caption{CLAR framework. During the training process, the reference and source samples are fed into our designed DDPM-based augmentation model to generate augmented data with new characteristics. These augmented data are further processed by cropping and resizing to build the contrastive loss. Meanwhile, the weight of each sample pair is computed by our devised adaptive weight algorithm and further is incorporated into the contrastive loss to enhance model performance.}
	\label{fig:Frame}
	\vspace{-0.3cm}
\end{figure}

%In this section, we present the diffusion model-based contrastive self-supervised learning framework for human activity recognition (CLAR). First, we illustrate an overview of the recognition framework. Next, we present the DDPM-based data augmentation method, which will generate augmented data with new characteristics. Finally, we illustrate the adaptive weight algorithm to compute weights of different positive sample pairs for the contrastive loss.

Here, we demonstrate the diffusion model-based contrastive learning framework for activity recognition (CLAR),
whose overview is illustrated in Figure~\ref{fig:Frame}. In this model, the source sample and the reference sample are first combined by our designed DDPM-based time series-specific augmentation method to generate the augmented sample. The two augmented samples derived from the same source sample are regarded as a positive pair. Then, the augmented samples are processed by the cropping and resizing operations for building the contrastive loss. During this procedure, the weights of the sample pairs are computed by our proposed adaptive weight algorithm, and further incorporated into the contrastive loss to enhance the model robustness.

\subsection{DDPM-based Data Augmentation Model}
\label{subsec:Augmented}

\begin{figure}[tb]
	\centering
	\includegraphics[width=0.40\textwidth]{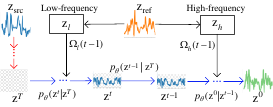}
	%\vspace{-0.3cm}
	\caption{The DDPM-based data augmented model. Red arrows $\color{red} \to $ indicate the forward diffusion, and blue ones $ \color{blue} \to $ refer to the reverse diffusion.}
	\label{fig:Diffusion}
	\vspace{-0.3cm}
\end{figure}

To generate augmented data with new motion habits, we design a DDPM-based data augmentation model, which can combine two samples from users with different habits to generate an augmented sample with the combined characteristics.
%------------------------------rpt-------------------
%To generate augmented data with new motion habits, we design a DDPM-based data augmentation model, which has the capability to blend two samples from users with distinct behaviors to produce an augmented sample with the combined characteristics.
Specifically, inspired by the superiority of DDPM on image and time series processing~\cite{2021WaveGrad, rombach2022high,2022BDDM,xiao2023imputation}, we introduce DDPM as the basic generative model.
We feed a prior derived from the source sample into the diffusion model and regard the reference sample as the condition imposed on the reverse diffusion (generative) process. 
However, for CSI data, directly imposing the condition on the generative process might cause data distortion. 
CSI data collected in the presence of activities, especially high-speed movements, will generate high frequency changes~\cite{wang2015understanding}. 
However, at the earlier generative steps, the generated latent variable is quite noised and flat, and directly injecting CSI data with a big fluctuation into this variable might make the data distorted. 
Therefore, to avoid this issue, unlike general conditional diffusion models which adopt direct imposition~\cite{rombach2022high,2022BDDM}, we decompose the condition (the reference sample) into the high-frequency part and low-frequency part, and then impose the former with increasing weights and the latter with decreasing weights on the generative process.
The reason is that at the earlier steps, the flat latent variable has the capacity of involving more low-frequency features with a small fluctuation, but not for high-frequency features with a big fluctuation. 
Also, the earlier steps may require more low-frequency features to capture global patterns, while the later steps may require more high-frequency features to capture local patterns~\cite{huang2023diffar}.

Figure~\ref{fig:Diffusion} presents the framework of the designed DDPM-based augmentation model.
In this model, we first use discrete wavelet transform~\cite{2016DWT} to extract the high-frequency features $\textbf{z}_h$ and low-frequency features $\textbf{z}_l$ from the reference sample. 
Then, we adopt the forward process to add noise into the source sample ${\textbf{z}_\text{src}}$ to form a prior ${\textbf{z}^{T}}$. The prior ${\textbf{z}^{T}}$ is fed into the reverse diffusion process to generate a clean CSI data through gradual denoising, i.e.,
${\textbf{z}^T}  \to \cdots \to {\textbf{z}^{t}} \to {\textbf{z}^{t - 1}}\to  \cdots \to {\textbf{z}^0}$.
During this denoising process, the features $\textbf{z}_h$ and $\textbf{z}_l$ are iteratively injected into the latent variable $\textbf{z}^{t}$ with varying weights $\Omega_h(t-1)$ and $\Omega_l(t-1)$, individually.
As a result, the generated (augmented) data $\textbf{z}^0$ contains the combined characteristics of both source and reference samples.
Assuming the source sample refers to CSI data performed by the user tending to draw a big circle and the reference sample for a small circle, the augmented data can be considered as the one for a middle circle.

%----Diffusion Autoencoders: Toward a Meaningful and Decodable Representation, cvpr 2022

%----Channel_State_Reconstruction_Using_Multilevel_Discrete_Wavelet_Transform_for_Improved_Fingerprinting-Based_Indoor_Localization, 
Concretely, we first extract the high-frequency features $\textbf{z}_h$ and low-frequency features $\textbf{z}_l$ using the high-pass filter $f_h$ and low-pass filter $f_l$, which are derived from the scaling functions and the corresponding wavelets~\cite{2016DWT}:
\begin{equation} \begin{aligned}
  & {\textbf{z}_h} = f_h(\textbf{z}_\text{ref}) ,   
  & {\textbf{z}_l} = f_l(\textbf{z}_\text{ref}).
\label{equ:concentrate}
\end{aligned} \end{equation}
%where $\textbf{z}_\text{ref}$ represents the original signal, $\textbf{z}_l$ and $\textbf{z}_h$ are the the low-pass and high-pass components of $\textbf{z}_\text{ref}$, respectively.
%In general, the high-frequency features $\textbf{z}_h$ contain more activity information with larger fluctuations and the reverse is true for the low-frequency features $\textbf{z}_l$.
Since the constant condition is improper for CSI data, we  introduce two monotonic functions $\Omega_h(t-1)$ and $\Omega_l(t-1)$ to adjust the weights of $\textbf{z}_h$ and $\textbf{z}_l$, individually. These two functions have a co-domain from 0 to 1 and can be obtained based on the exponential equation:
\begin{equation} \begin{aligned}
\Omega_h(t-1) =  N_h e^{-\lambda (t-1)}, 
\label{equ:decearsing}
\end{aligned} \end{equation}
\begin{equation} \begin{aligned}
\Omega_l(t-1) =  N_l e^{-\lambda (T-t)},
\label{equ:growth}
\end{aligned} \end{equation}
where $\lambda_h$ and $\lambda_l$ are the hyper-parameters representing their exponential decay and growth constants, respectively, $N_h$ and $N_l$ are the initial quantities, and $T$ is the number of the diffusion steps.
As step $t$ decreases and reverse sampling continues, the weight of the high-frequency feature, $\Omega_h(t-1)$, will increase and the weight of the low-frequency feature, $\Omega_h(t-1)$, will decrease. In this way, the characteristics of the reference sample can be injected into the source sample without data distortion.

%However, in the earlier steps (i.e., the larger $t$), simply injecting the condition $\textbf{z}_h$ into the denoising network might cause the deviation problem. This is because in the smaller steps the generated latent variable $\textbf{z}^{t-1}$ is quit noised and flat, and exerting the condition $\textbf{z}_h$ with larger fluctuations might distort the generated data. On the other hand, in the later steps, the generated values become cleaner, and imposing the condition $\textbf{z}_h$ with larger weights based on the the warping path will not cause serious distortion. On the contrary, low-frequency features $\textbf{z}_l$ with smaller fluctuations can well be imposed on the generated data in the earlier steps, but not in the latter steps. Also, the earlier steps may require the low-frequency features to capture global patterns, while the later steps may require the high-frequency features to capture local patterns~\cite{huang2023diffar}. Hence, we propose a novel varying conditioner to dynamically adjust the weights of both conditions $\textbf{z}_h$ and $\textbf{z}_l$. In other words, 

Before imposing the high-frequency and low-frequency features on the generative process, we first adopt the forward process (Equation~\ref{equ:forward}) to compute $\textbf{z}_h^{t - 1}$ and $\textbf{z}_l^{t - 1}$ from $\textbf{z}_h$ and $\textbf{z}_l$, respectively:
\begin{equation} \begin{aligned}
    &\textbf{z}_h^{t-1} \sim q(\textbf{z}_h^{t-1} \lvert \textbf{z}_h),
    \\
    &\textbf{z}_l^{t-1} \sim q(\textbf{z}_l^{t-1} \lvert \textbf{z}_l).
\label{equ:forwardSrcRef}
\end{aligned} \end{equation}
Then, we utilize the reverse process (Equation~\ref{equ:reverse}) to compute latent variable $\hat{\textbf{z}}^{t-1}$ from $\textbf{z}^t$:
\begin{equation} \begin{aligned}
    &\hat{\textbf{z}}^{t-1} \sim p_\theta(\hat{\textbf{z}}^{t-1} \lvert \textbf{z}^t).
\label{equ:reverseNoise}
\end{aligned} \end{equation}
As a result, considering the weights $\Omega_h(t-1)$ and $\Omega_l(t-1)$, we impose the conditions, $\textbf{z}_h$ and $\textbf{z}_l$, on the reverse diffusion process ${p_\theta }\left( {{\textbf{z}^{t - 1}}|{\textbf{z}^t}} \right)$ by  matching $f_l(\hat{\textbf{z}}^{t-1})$ of $\hat{\textbf{z}}^{t-1}$ with that of $f_h(\textbf{z}_h^{t - 1})$ and $f_l(\textbf{z}_l^{t - 1}))$ as follows:
\begin{equation}
\begin{split}
    \textbf{z}^{t-1}= &\hat{\textbf{z}}^{t-1} + \Omega_h(t-1) \left(f_h\big(\sigma (\hat{\textbf{z}}^{t-1},\textbf{z}_h^{t - 1})\big)-f_h(\hat{\textbf{z}}^{t-1})\right) \\
    &+ \Omega_l(t-1) \left(f_l\big(\sigma (\hat{\textbf{z}}^{t-1},\textbf{z}_l^{t - 1})\big)-f_l(\hat{\textbf{z}}^{t-1})\right),
\end{split}
\label{equ:OppFinal}
\end{equation}
where $\sigma$ is a aggregation function which concatenates the two variables based on the warping path.
Here, the warping path, produced using dynamic time warping~\cite{Dynamic2017,Dynamic2020}, maps the elements of two data sequences to minimize the distance between them.
%------------------------------rpt-------------------
%%The warping path, generated through the use of dynamic time warping~\cite{Dynamic2017,Dynamic2020}, is employed to map the elements of two data sequences with the goal of minimizing the distance between them.
We adopt the warping path, instead of the default shortest path, to concatenate them because the warping path can more appropriately keep the shape of the waveforms~\cite{xiao2022self}.
%------------------------------rpt-------------------
%We adopt the warping path, as opposed to the default shortest path, to concatenate them because the warping path can more appropriately keep the shape of the waveforms~\cite{xiao2022self}.

The matching operation by Equation~\ref{equ:OppFinal} ensures that the conditions $\textbf{z}_h$ and $\textbf{z}_l$ are exerted into the reserve process with varying weights.
In the early diffusion steps, as the the latent variable is quite noised and flat, to avoid the data distortion, the high-frequency feature $\textbf{z}_h$ is assigned a smaller weight and the low-frequency feature $\textbf{z}_l$ for a bigger weight. 
On the other hand, the later steps consider more high-frequency features and less low-frequency feature to involve the activity characteristics.
In this way, as the features from the reference samples are injected into the latent variable in the generative process and the generative process starts with the prior derived from the source sample, the generated (augmented) data can possess compromised characteristics of them. Hence, the augmented data can be considered to be the one collected from another user with the different habit.
Both augmented samples and source samples will be used for model training.

\subsection{Adaptive Weighting}
\label{subsec:Weight}

In contrastive learning, one critical question is how to construct positive pairs~\cite{SimCLR2020,gao2021simcse}.
Cropping is a commonly used way to extract views for building positive sample pairs~\cite{peng2022crafting, What2020}. 
In other words, the two crops derived from the same sample are regarded as a positive pair, and the crops from different samples as a negative pair.
For each activity data, we also adopt cropping operations to extract two views (samples) from the same activity data to form a positive pair.
While, for CSI data, the clues provided by different positive pairs should be various in learning data representation. 
For some activities, there may be a pause in the action. For example, for drawing X, a pause can occur between the two strokes, and for lying down, it can occur between sitting and lying.
Hence, some positive pairs extracted by cropping operations might contain more activity data, while others might include more pause data. 
Correspondingly, the positive pairs containing more activity data should provide more clues for learning data representation, and vice versa. 
%------------------------------rpt-------------------
%Correspondingly, the positive pairs containing more action data should provide more clues for obtaining data representations, and vice versa. 
An example of drawing X is presented in Figure~\ref{fig:pauseExample}, where the positive pair ($x_3$, $x_4$) should play a  more important role for model training than ($x_1$, $x_2$), because the former contains more activity data.
%------------------------------rpt-------------------
%An example of drawing X is presented in Figure~\ref{fig:pauseExample}, where the positive pair ($x_3$, $x_4$) is expected to have a greater impact on model training compared to ($x_1$, $x_2$), because the former contains more activity data.

Toward this goal, we propose an adaptive weight algorithm to adjust the importance of positive pairs for model training by assigning various weights to different positive pairs.
This algorithm first computes a response map that can reflect the amount of activity data in the positive pair and then computes the weights based on the response map for constructing the contrastive loss.

%First, the continuous data streams are split into overlapping windows using a sliding window, each with length w, where the sliding step is 1. Second, the state label of each window is inferred using the state inference model.

Concretely, to compute the response map, 
we first select a template $w^T$ with length $H$ from the CSI sequence in the absence of activity data, called \textit{static template}. To avoid selection bias, we choose multiple static templates. 
Then, for each sample from positive pairs, we split it into overlapping windows using a sliding window, each with length $H$, where the sliding step is 1.
For window $l$ extracted from sample $x_i$, we adopt a response score to reflect the amount of containing activity data:
% CSI data in the presence of activities
\begin{equation} \begin{aligned}
S_{l} = \frac{1}{K} \sum_{k=1}^{K} \text{DTW}(w_l, w_k^T),
\label{equ:dwtSimilar}
\end{aligned} \end{equation}
where $K$ is the number of selected static templates, and $\text{DTW}(w_l, w^T)$ denotes the Dynamic Time Warping (DTW)~\cite{Dynamic2017,Dynamic2020} distance between the $w_l$ and $w^T$.
The bigger distance between $w_l$ and $w^T$ indicates that $w_l$ is more different from static template $w^T$, i.e., $w_l$ contains more activity data. Therefore, response score $S_{l}$ reflects the amount of activity data in window $l$. 
The response scores of the windows in $x_i$  are merged to form the response map of this sample.

After obtaining response maps, we calculate the weights of  sample $x_i$ for model training:
\begin{equation} \begin{aligned}
{W_{i}} = {\left( {\frac{1}{N_w} \sum_{k = 1}^{N_w} {\rm{I}}\left(S_k,\sigma _s \right)} \right)^\alpha },
\label{equ:weight}
\end{aligned} \end{equation}
where $\alpha$ denotes the power which controls the scale of weights,  $N_w$ refers to the number of the windows extracted from $x_i$, and ${\rm{I}}(,)$ is the indicator of the presence of activity data,
and is defined as:
\begin{equation} \begin{aligned}
{\rm{I}}({S_k},{\sigma_s}) = \left\{ {\begin{array}{*{20}{l}}
{\;1,}&{{\rm{if  }}{S_k} > {\sigma _s}}\\
{\;0,}&{{\rm{otherwise}}}
\end{array}} \right..
\end{aligned} \end{equation} 
Here ${\sigma _s}$ is a threshold to determine whether this window is regarded as data in the presence of activities. ${\sigma _s}$ can be set to the average of the response scores, i.e., ${\sigma _s} = {{(\sum\nolimits_{k = 1}^{N_w} {S_{k}} )} \mathord{\left/
 {\vphantom {{(\sum\nolimits_{i = 1}^{N_w} {S_{k}} )} N_w}} \right.\kern-\nulldelimiterspace} N_w}$.
Further, for a positive pair ($x_i, x_j$), its weight is the aggregation of the weights of both samples:
\begin{equation} \begin{aligned}
W_{(i,j)} = \text{Aggregate}(W_i, W_j),
\label{equ:Weight}
\end{aligned} \end{equation}
where $\text{Aggregate}(,)$ sums the two items.
This weight suggests the amount of containing CSI data in the presence of activities. Hence, the positive pairs with a larger weight contains more clues and should play a more significant role in model training.

%First, we randomly crop the augmented samples to get a pair of samples (${x_1}$,${x_2}$), and then take a section of static waveform t and use Dynamic Time Warping (DTW) ~\cite{Dynamic2017}~\cite{Dynamic2020} to slide on ${x_1}$ and ${x_2}$ to calculate the similarity, and obtain two response maps. Therefore, the response map represents the possibility that a pair of samples are in a static state. If the similarity calculated by sliding is higher, the probability of static waveform is higher, and the lower the similarity is, the lower the probability of static waveform is.

%Then we calculate the weight according to the response map by Equation~\ref{equ:weight}. 
%If the calculated similarity of each sliding window is greater than the threshold, the indicator function value is 1, otherwise it is 0.
%Therefore, we propose an adaptive weighting strategy to determine the importance of each training sample by considering the proportion of indicator function of 1 in the response map. 
%If the proportion of the indicator function is 1, the higher the proportion of the sample to the static waveform, the lower the weight in the sample training. On the contrary, the lower the proportion, the higher the weight.
%The weight are defined as follows:

\subsection{Overall Model}
\label{subsec:Overall}

Taking the augmented data and adaptive weights into account, we formulate the loss function as follows:
\begin{equation} \begin{aligned}
     \mathcal{L}_{i,j}^{aug} =  - \log \frac{{\exp \left( {{{{W_{(i,j)}} * sim\left( {{{\hat{\textbf{z} }}_i},{{\hat{\textbf{z} }}_j}} \right)} \mathord{\left/
 {\vphantom {{{W_{i,j}^k} * sim\left( {{{\hat{\textbf{z} }}_i},{{\hat{\textbf{z} }}_j}} \right)} \tau }} \right.
 \kern-\nulldelimiterspace} \tau }} \right)}}{{\sum\nolimits_{k = 1}^{k = 2M} {{{\rm I}_{k \ne i}}\exp \left( {sim{{\left( {{{\hat{\textbf{z} }}_i},{{\hat{\textbf{z} }}_k}} \right)} \mathord{\left/
 {\vphantom {{\left( {{{\hat{\textbf{z} }}_i},{{\hat{\textbf{z} }}_k}} \right)} \tau }} \right.
 \kern-\nulldelimiterspace} \tau }} \right)} }},
\label{equ:augLoss} 
\end{aligned} \end{equation}
where $M$ is the length of the minibatch, $\tau$ is a temperature hyper-parameter scaling the distribution of distances. $\hat{\textbf{z}}_i$ and $\hat{\textbf{z}}_j$, which form a positive pair, are two embeddings that are extracted from the two augmented samples derived from the same source sample, and  $\hat{\textbf{z}}_i$ and $\hat{\textbf{z}}_k$, which form a negative pair, are derived from the different source samples.

Further, we also adopt the original data without the process of our designed augmentation model to build the contrastive loss to capture the characteristics of the original training data. This loss is defined as:
\begin{equation} \begin{aligned}
     \mathcal{L}_{i,j}^{ori} =  - \log \frac{{\exp \left( {{{{W_{(i,j)}} * sim\left( {{\textbf{z}_i},{\textbf{z}_j}} \right)} \mathord{\left/
 {\vphantom {{{W_{i,j}^k} * sim\left( {{\textbf{z}_i},{\textbf{z}_j}} \right)} \tau }} \right.
 \kern-\nulldelimiterspace} \tau }} \right)}}{{\sum\nolimits_{k = 1}^{k = 2M} {{{\rm I}_{k \ne i}}\exp \left( {sim{{\left( {{\textbf{z}_i},{\textbf{z}_k}} \right)} \mathord{\left/
 {\vphantom {{\left( {{\textbf{z}_i},{\textbf{z}_k}} \right)} \tau }} \right.
 \kern-\nulldelimiterspace} \tau }} \right)} }},
\label{equ:oriLoss} 
\end{aligned} \end{equation}
where $\textbf{z}_i$, $\textbf{z}_j$ and $\textbf{z}_k$ are the embeddings of the original samples without being processed by our augmentation model, and the other parameters are the same as Eqation~\ref{equ:augLoss}.
As a result, the overall loss is the sum of them:
\begin{equation} \begin{aligned}
     \mathcal{L}_{i,j}^{all} =  \mathcal{L}_{i,j}^{aug} + \mathcal{L}_{i,j}^{ori}.
\label{equ:finalLoss} 
\end{aligned} \end{equation}

After obtaining the trained model, it is used to extract representations of activity samples. Further, a linear classifier is adopted to classify the representations into corresponding activity categories.

\section{Experimental Evaluation}
\label{sec:experiment1}

In this section, we evaluate the effectiveness of the proposed CLAR by comparing it with several baselines with different techniques. Also, we conduct ablation studies and inspect the role of the augmentation model and labeled data size. The data and code are available online\footnote{https://github.com/ChunjingXiao/CLAR}.
%------------------------------rpt-------------------
%In this section, we investigate the efficacy of the proposed CLAR by comparing it with several baselines that utilize different techniques. Also, we inspect the role of our designed components, the augmentation model and labeled data size. The data and code are available online\footnote{https://github.com/ChunjingXiao/CLAR}.

\subsection{Experiment Setup}
\label{subsec:setup}
For the evaluation, we conduct experiments on two WiFi CSI-based behavior recognition datasets.
%------------------------------rpt-------------------
%For the evaluation, we carry out experiments on two datasets for action recognition based on WiFi CSI.
\textbf{SignFi data}~\cite{2018signfi} consists of 1,250 CSI sequences, each of which represents a sign language gesture.
These activities are performed by 4 users with each activity repeated for 10 times.
\textbf{DeepSeg data}~\cite{DeepSeg2021} is composed of 1,500 human activities from 5 users with various shapes and ages.
%For these experiments, We will use 60\% of all data as the training set, 20\% as validation set and the rest as the test set.
For these experiments, we use 80\% of all data as the training set and the rest as the test set. 
Here, the test data is always unseen during the training process.
Since collecting numerous labeled data is time-consuming and labor-intensive, we assume that the labeled data is inadequate.
We select 30\% and 20\% of data as labeled data for fine-tuning the classifier on SignFi and DeepSeg, respectively, since they can obtain the expected performance at a lower cost.

For the selection of source and reference samples, if they are labeled, we select two samples from the same activity category as the source and reference ones.
If they are unlabeled, for a source sample, we randomly select one sample from its top 10 most similar samples as the reference one. Here, we adopt DTW to compute the similarity degree between samples.
For the DDPM-based augmentation model, we use 1,000 diffusion steps considering both efficiency and effectiveness.
During the optimization process of CLAR, the learning rate and the mini-batch size are set to 0.0001 and 50, respectively.
The hyper-parameters $\alpha$ in Equation~\ref{equ:weight} and $\tau$ in Equation~\ref{equ:augLoss} are set 0.5 and 0.1, individually, for both datasets. For all the following experiments, the accuracy and F1-score are adopted as metrics for performance comparison.
%------------------------------rpt-------------------
%The hyper-parameters $\alpha$ in Equation~\ref{equ:weight} and $\tau$ in Equation~\ref{equ:augLoss} are set 0.5 and 0.1, individually.

%We selectthe biorthogonal wavelet~\cite{AntoniniImage} with a two-layer structure to perform DWT.

%The former refers to the ratio of samples whose categories can be properly predicted. The latter means the combination of Recall and Precision with the same weight.

%We use VGG16 as the feature extraction network. Because the size of CSI data is 200x30x3, we replace the size and side of MaxPooling on the last layer from (2, 2) to (3, 1), and remove the full connection layer. For nonlinear projection layer, we use two-layer MLP. For the data enhancement function, we set the color jitter intensity parameter to 0.5, the random flip probability to 0.5, and use the same parameter settings for the SignFi and DeepSeg data sets.

\subsection{Baselines}
\label{subsec:baselines}

To prove the effectiveness and superiority of the proposed model, we choose activity recognition methods with different technologies as the baselines,
%------------------------------rpt-------------------
%To demonstrate the efficacy of CLAR, we choose activity recognition approaches that utilize various technologies as baseline methods,
including supervised methods~\cite{huang2023diffar, 2024ViTHAR}, few-shot approaches~\cite{Across2022, 2024CDFi, 2024Metaformer} and self-supervised contrastive learning models~\cite{2022ColloSSL, yang2023autofi, 2024DFWS}:

\begin{itemize} %[leftmargin=*]

    \item \textbf{DiffAR}~\cite{huang2023diffar}: A temporal-augmented supervised learning method for human activity recognition. It presents a novel conditional diffusion model to produce augmented CSI, effectively addressing the problem of missing values.
    %An activity recognition model based on GANs utilizing WiFi CSI data. This approach presents a novel complementary generator and enhances the performance of activity recognition by optimizing the outputs and loss functions of the discriminator.
   \item \textbf{ViTHAR}~\cite{2024ViTHAR}: 
   A Transformer-based human activity recognition model. This method tries to apply five different Vision Transformers to WiFi CSI-based activity recognition.
   %A supervised learning-based human activity recognition method using a Vision Transformer. This method applies five different Vision Transformers on two WiFi CSI datasets for human activity recognition.
   % A supervised learning method based on GANs that includes manifold regularization. This model shows significant advantages over other semi-supervised methods that use GANs for image classification tasks.
    %\item \textbf{ManiGAN}~\cite{Bruno2018Semi}: A semi-supervised learning method based on GANs that includes manifold regularization. This model shows significant advantages over other semi-supervised methods that use GANs for image classification tasks

    %\item \textbf{CsiGAN}~\cite{2019CsiGAN}: An activity recognition model based on GANs utilizing WiFi CSI data. This approach presents a novel complementary generator and enhances the performance of activity recognition by optimizing the outputs and loss functions of the discriminator.

    %\item \textbf{RF-Net}~\cite{2021RF-Net}: A unified meta-learning framework for RF-enabled one-shot activity recognition. It delivers the capability of being adaptive to new environments with very few labeled data.
    %------------------------------rpt-------------------
    %\item \textbf{RF-Net}~\cite{2021RF-Net}: A comprehensive meta-learning method designed for one-shot activity recognition. This approach provides adaptability to new environments with minimal labeled data.
    
    \item \textbf{MetaAct}~\cite{Across2022}: A meta learning-based adaptable activity recognition model. This approach is specifically designed for recognizing activities across scenes and categories using WiFi CSI.
    %------------------------------rpt-------------------
    %\item \textbf{MetaAct}~\cite{Across2022}: A few-shot learning-based adaptable action recognition framework. The method is particularly tailored for action recognition across different settings and classes with CSI.
   \item \textbf{CDFi}~\cite{2024CDFi}: A few-shot learning-based cross-domain WiFi action recognition method. This method transfers the knowledge from the source domain to the target environment for cross-domain activity recognition.
    %A comprehensive meta-learning method designed for one-shot activity recognition. This approach provides adaptability to new environments with minimal labeled data.    
    \item  \textbf{MetaFormer}~\cite{2024Metaformer}: a few-shot learning-based WiFi sensing system. This system can effectively recognize actions from unseen domains with only one labelled target sample per category.      
   % A comprehensive meta-learning method designed for one-shot activity recognition. This approach provides adaptability to new environments with minimal labeled data.
    %\item \textbf{MultiSSL}~\cite{2019MulSSL}: A self-supervised learning method for human activity recognition. MultiSSL learns accelerometer representations by training a temporal convolutional neural network to recognize the transformations applied to the raw input signal.
    %------------------------------rpt-------------------
    %\item \textbf{MultiSSL}~\cite{2019MulSSL}: A self-supervised learning method for action recognition. MultiSSL trains a temporal convolutional network to identify the transformations that are utilized in the raw signal to learn representations of accelerometers.

    \item  \textbf{ColloSSL}~\cite{2022ColloSSL}: A collaborative self-supervised learning-based human activity recognition approach. It leverages unlabeled data collected from multiple devices to learn high-quality features of the data.
 
    \item \textbf{AutoFi}~\cite{yang2023autofi}: A self-supervised learning activity recognition model using WiFi CSI. AutoFi fully utilizes unlabeled low-quality CSI samples to learn the knowledge, which is further transferred to speciﬁc tasks. 

   \item  \textbf{DFWS}~\cite{2024DFWS}: A contrastive learning-based human activity recognition method. DFWS maximizes the mutual information between the features and the corresponding samples to obtain features of unlabeled samples.
  
    %A self-supervised learning method for action recognition. MultiSSL trains a temporal convolutional network to identify the transformations that are utilized in the raw signal to learn representations of accelerometers.
    
    %\item \textbf{ColloSSL}~\cite{2022ColloSSL}: A self-supervised learning method for learning from unlabeled inertial sensor data collected from multiple devices worn by the user. ColloSSL leverages natural transformations in the sensor datasets collected from multiple devices to perform contrastive learning, and learns a robust feature extractor for downstream HAR classiffcation tasks.
        
	%This method is to randomly change an image to obtain a pair of augmented images $x_{i}$ and $x_{j}$ of the image $x$.
	%And for this pair of augmented images, the encoder is used to obtain the image representation, and then a projection network is used to obtain the image representation z, the purpose of which is to maximize the similarity between the two representations of the same image, $z_{i}$ and $z_{j}$.
	
\end{itemize}

\begin{figure}[tb]
	\centering
	\includegraphics[width=0.40\textwidth]{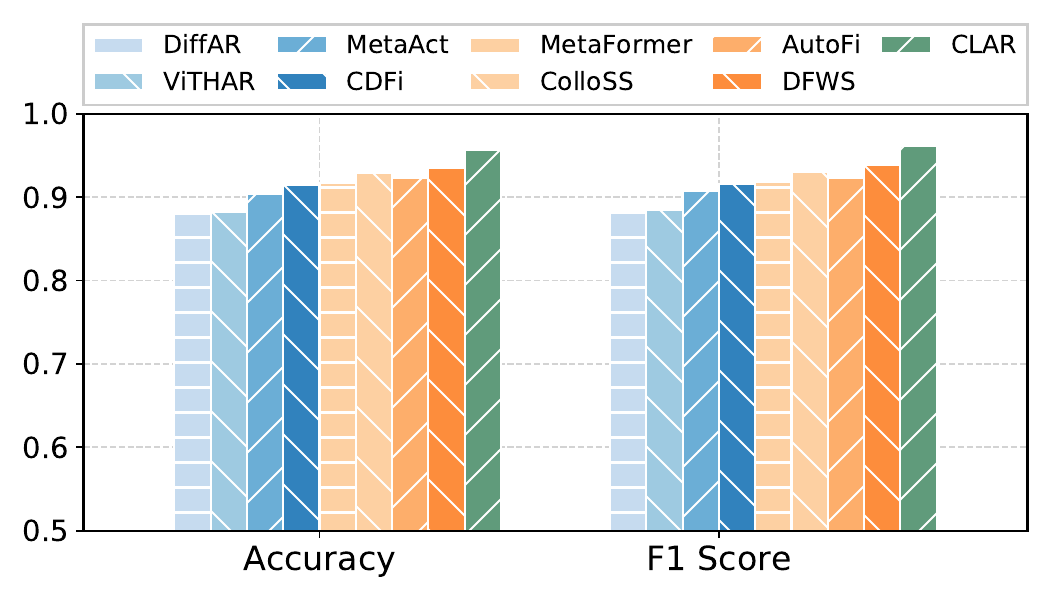}
	%\vspace{-0.3cm}
	\caption{The activity recognition performance for SignFi data.}
	\label{fig:SignFi}
	\vspace{-0.3cm}
\end{figure}

\begin{figure}[tb]
	\centering
	\includegraphics[width=0.40\textwidth]{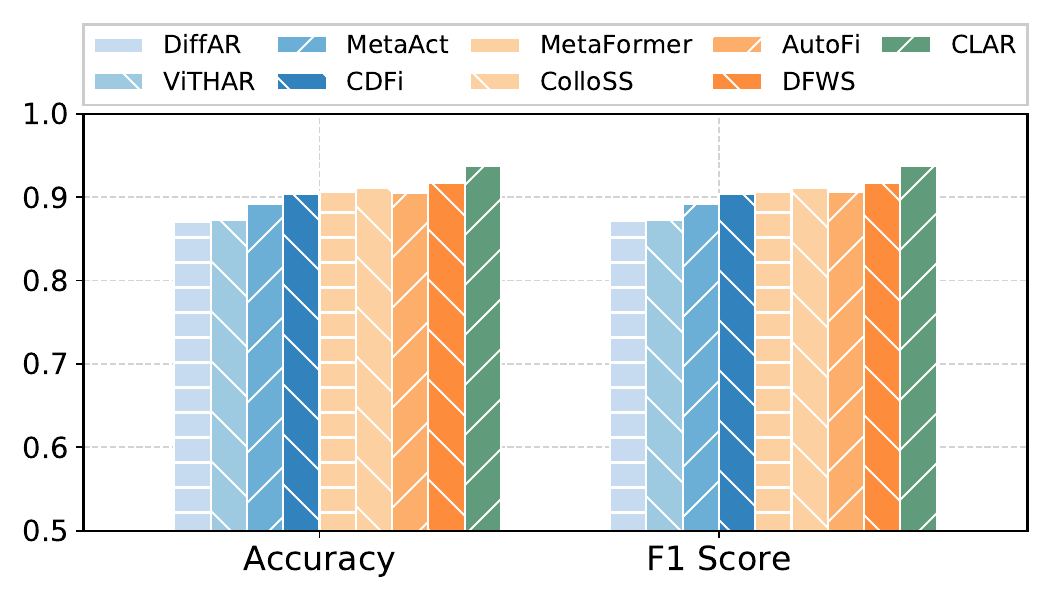}
	%\vspace{-0.3cm}
	\caption{The activity recognition performance for DeepSeg data.}
	\label{fig:DeepSeg}
	\vspace{-0.3cm}
\end{figure}

\subsection{Recognition Performance Comparison}
\label{subsec:Performance}

Figure~\ref{fig:SignFi} and~\ref{fig:DeepSeg} report the results of our model and the baseline models across the two datasets: SignFi and DeepSeg. From these results, we have the following observations.
First, our model CLAR consistently yields better performance on the two datasets.
For example, compared to the contastive learning method DFWS, CLAR exhibits improvements of around 2.2\% and 2.0\% on the SignFi and DeepSeg datasets, respectively. CLAR achieves more distinct improvements on SingFi. The reason is that there are more activity categories on SingFi, meaning fewer labeled samples per category. The limited labeled samples lead to inferior performance for the baselines. However, by generating augmented data and taking advantage of unlabeled data, our method CLAR can efficiently address this issue and achieve higher performance.

Second, few shot learning-based methods (e.g., MetaAct, CDFi and MetaFormer), exceed the two supervised baselines, DiffAR and ViTHAR. Since few shot learning-based methods are designed for the scenario with several labeled samples. Therefore, under the environment with limited labeled data, the few shot learning-based models can obtain better performance than the supervised models, which generally require a given number of labeled samples to obtain the expected performance.

Third, self-supervised models outperform the other baselines. In particular, ColloSSL, which is designed for non-WiFi signal, also achieves relatively good performance, compared with the supervised models. This indicates that the self-supervised techniques can effectively benefit WiFi CSI-based human activity recognition, especially for the scenario with limited training data. 
However, by incorporating our designed augmentation model and adaptive weight algorithm, our model CLAR significantly outperforms these baselines.

\begin{table}[ht]
%\vspace{-0.2cm}
\caption{The performance comparison using the different classifiers.} % Title of Table
\label{table:classifier}
%\vspace{-0.3cm}
\centering
\begin{tabular}{p{0.9cm}<{\centering}|p{1.05cm}<{\centering}|p{1.1cm}<{\centering} p{1.1cm}<{\centering} p{1.1cm}<{\centering} p{1.1cm}<{\centering}}
\toprule
 \multirow{2}{*}{Classfier} & \multirow{2}{*}{Method} & \multicolumn{2}{c}{\text { SignFi }} & \multicolumn{2}{c} {\text{DeepSeg }} \\ 
 \cmidrule(lr){3-4} \cmidrule(lr){5-6}
 & &  \centering\text{Accuracy} & \centering{F1 Score} &  \centering\text{Accuracy} & \text \centering{F1 Score} \\  
                           
\midrule
\multirow{4}{*}{Linear}   &  ColloSSL  &  92.93   & 92.97	 & 91.04	& 91.12    \\
                          &  AutoFi    &	92.25   & 92.29	   & 90.53	  &	90.61    \\
                          &  DFWs    &	93.50   & 93.95	   & 91.65	  &	91.69     \\
                          &  CLAR      &	\textbf{95.70}   & \textbf{96.10}	   & \textbf{93.72}	  &\textbf{	93.74}\\
\midrule										
\multirow{4}{*}{Softmax}  &  ColloSSL  &  92.89   & 92.91	   & 91.01	  &	91.03\\
                          &  AutoFi    &	92.10   & 92.20	   & 90.51	  & 90.54  \\
                          &  DFWs    &	93.10   & 93.71	   & 91.59	  & 91.63 \\
                          &  CLAR      &	\textbf{95.13}   & \textbf{95.16}	   & \textbf{92.83}	  &\textbf{92.85}\\
\bottomrule
\end{tabular}
\vspace{-0.2cm}
\end{table}

Like general contrastive learning models, our model also aims at extracting high-quality representations. To investigate the effectiveness of the extracted representations, we here compare the recognition performance when using the same classifier with the three contrastive learning-based baselines: ColloSSL, AutoFi and DFWS. We observe the performance when utilizing two widely used classifiers for contrastive learning: the linear classifier and softmax classifier. The results are presented in Table~\ref{table:classifier}.
As shown, for both linear and softmax classifiers, our model CLAR consistently outperforms these three baselines. Since they adopt the same classifier, the performance improvement should be attributed to the better representations obtained by our model. This indicates that our model can learn efficient representations regardless of the type of classifiers.

\begin{figure}[tb]
	\centering
	\includegraphics[width=0.48\textwidth]{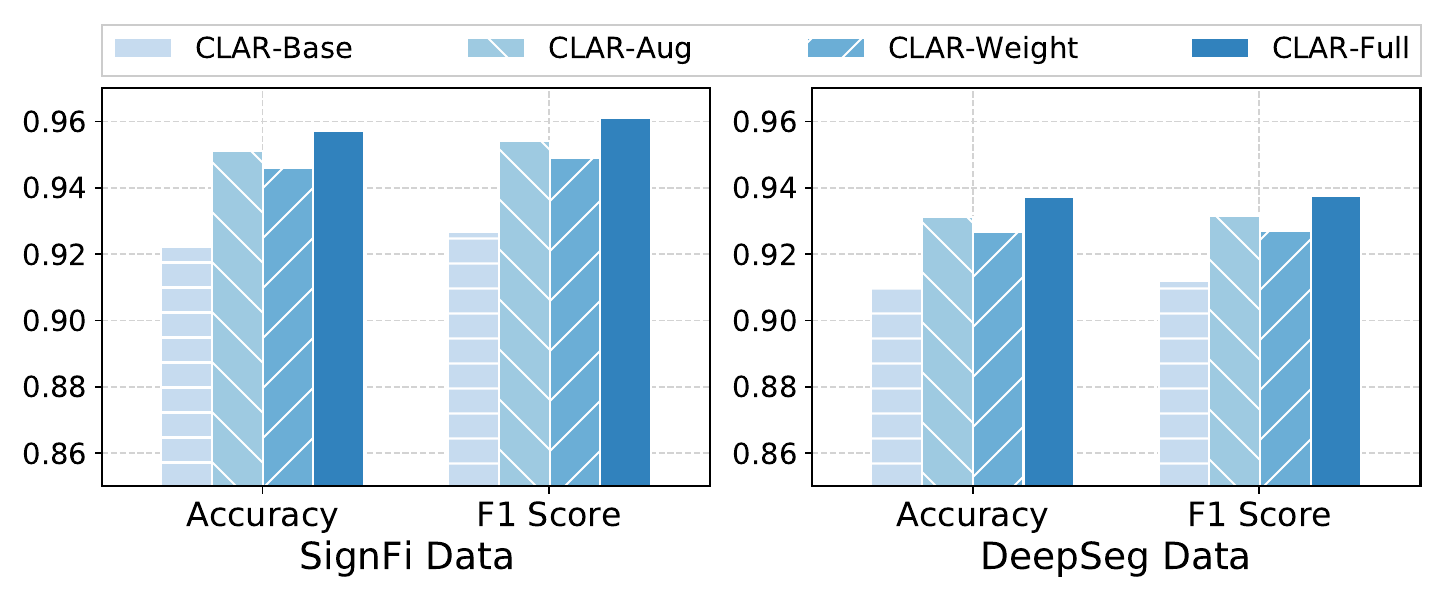}
	%\vspace{-0.3cm}
	\caption{The performance with different design choices for SignFi data and DeepSeg data.}
	\label{fig:ablation}
	\vspace{-0.2cm}
\end{figure}

\subsection{Ablation Study}
\label{subsec:Ablation Study}

Here we investigate the contribution of the two important components in CLAR, i.e., the augmentation model and adaptive weight algorithm. 
Specifically, we investigate the role of different components by considering the following variants of our model:
%Now we focus on studying the contributions of two essential components in our model, i.e.,
%We evaluate the two components¡¯ effectiveness by measuring the performance of the variant when one of the components is disabled.
%Specifically, we study the role of different components by considering following variants of our modle:
	(1)  \textit{CLAR-Base} is the basic contrastive learning framework that removes the DDPM-based augmentation model and the adaptive weight algorithm.
	%\item  \textbf{CLAR-Auxi} is the basic SimCLR model that incorporates an auxiliary task for the data enhancement method for time series
	(2)  \textit{CLAR-Aug} is the contrastive learning framework with the DDPM-based augmentation model but without the adaptive weight algorithm.
	(3)  \textit{CLAR-Weight} is the contrastive learning framework with the adaptive weight algorithm but without the DDPM-based augmentation model.
	(4)  \textit{CLAR-Full} is our proposed model fully incorporating all the components.
    %------------------------------rpt-------------------
	%(4)  \textit{CLAR-Full} is our proposed model fully involving these two modules.

%\begin{table}[ht]
%	%\vspace{-0.2cm}
%	\caption{The performance when incorporating different
%		components} % Title of Table
%	\label{table:choice}
	%\vspace{-0.3cm}
%	\centering
%	\begin{tabular}{p{0.95cm}|p{1.20cm}|p{0.80cm}<{\centering} p{0.80cm}<{\centering} p{0.80cm}<{\centering} %p{0.80cm}<{\centering} }
%		\toprule
%		&  DataSet   & CLAR-Base     & CLAR-Aug     & CLAR-Weight    & CLAR-Full                 \\
%		\midrule
%		\multirow{2}{*}{SignFi}   &  Accuracy  & 0.900 & 0.944 & 0.948 &\textbf{0.952}\\
%		                         &  F1 Score  & 0.902 & 0.947 & 0.949 &\textbf{0.956}\\
%		
%		\midrule										
%		\multirow{2}{*}{DeepSeg}  &  Accuracy   &	 0.910	 &	 0.926   & 	0.928  	   &\textbf{0.930}\\
%		                         &  F1 Score   &	0.891	 &	0.907    & 	0.909 	   &\textbf{0.915}\\
%		\bottomrule
%	\end{tabular}
%	\vspace{-0.3cm}
%\end{table}

The experimental results using SignFi data and DeepSeg data are presented in Figure~\ref{fig:ablation}. We summarize the observations from this figure as follows. First, CLAR-Full performs the best, while CLAR-Base is the worst model, which implies that the main components we proposed can significantly improve the recognition performance. 
%------------------------------rpt-------------------
%The outcomes of the experiments conducted with SignFi and DeepSeg are depicted in Figure~\ref{fig:ablation}. We can summarize the findings from this figure. First, CLAR-Full performs the best while CLAR-Base is the worst performing method, suggesting that our main modules can substantially enhance the identification results. 
Second, when incorporating the DDPM-based augmentation model, CLAR-Aug obtains better results than CLAR-Base. This is because the limited samples are augmented by our designed method, which can benefit the model in improving the generalization capacity. 
%------------------------------rpt-------------------
%Second, CLAR-Aug outperforms CLAR-Base when incorporating the DDPM-based augmentation model. This is likely due to the fact that our designed method augments the limited samples, which benefits facilitating the generalization capacity.
Third, CLAR-Weight outperforms CLAR-Base by a certain margin. The results prove the motivation of our model, i.e., introducing the adaptive weights can enable the model to capture more characteristics of activity data and further significantly enhance recognition performance.
%------------------------------rpt-------------------
%Third, CLAR-Weight outperforms CLAR-Base by a certain margin. These results validate the motivation behind our method, which is to introduce adaptive weights that enable the model to capture more features of activity data and remarkably boost recognition results.

\subsection{Role of the Augmentation Model}
\label{subsec:augmentMethod}

The analyses in the previous section suggest our designed augmentation model can effectively contribute to the performance improvement. Here, we further inspect the efficacy of the augmented data when applying them to other activity recognition models. 
%------------------------------rpt-------------------
%The above analysis indicates our designed augmentation model can effectively contribute to the performance improvement. In this section, we will further inspect the efficacy of augmented data when using them with other activity recognition models.
%----
We select four baseline approaches that are specially designed for WiFi CSI-based activity recognition under cross scenes:     \textit{MetaAct}~\cite{Across2022},
\textit{CDFi}~\cite{2024CDFi},
\textit{MetaFormer}~\cite{2024Metaformer},
and \textit{AutoFi}~\cite{yang2023autofi}. 
We evaluate the model performance with/without the augmented data for model training, named as \textit{one-with-aug}/\textit{one-non-aug}. To inspect the generalization capacity, we conduct these experiments under the left-out scene, i.e. the data of one user are extracted as the test data, and others as the training data.
%------------------------------rpt-------------------
%We assess the model's performance with/without augmented data, referred to as \textit{one-with-aug}/\textit{one-non-aug}. To inspect the generalization capacity, we conduct these experiments under the left-out scene, i.e. the data of one user are extracted as the test, and the rest as the training.

\begin{figure}[tb]
	\centering
	\includegraphics[width=0.48\textwidth]{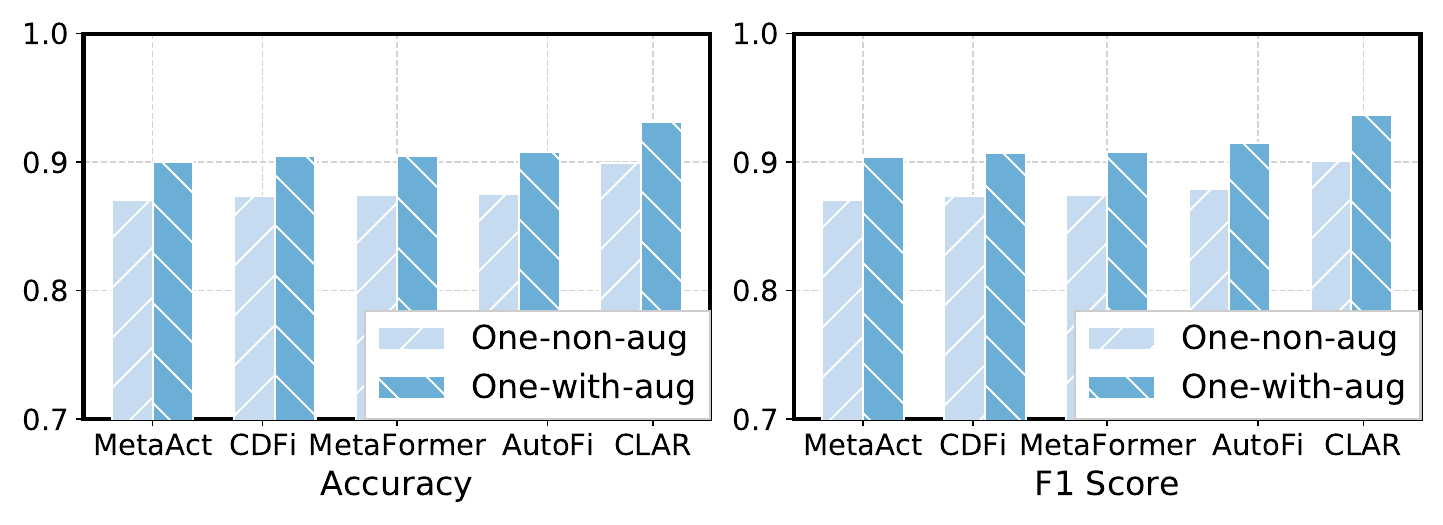}
	%\vspace{-0.3cm}
	\caption{The performance with/without the augmented data for SignFi data.}
	\label{fig:SignFiAug}
	\vspace{-0.2cm}
\end{figure}

\begin{figure}[tb]
	\centering
	\includegraphics[width=0.48\textwidth]{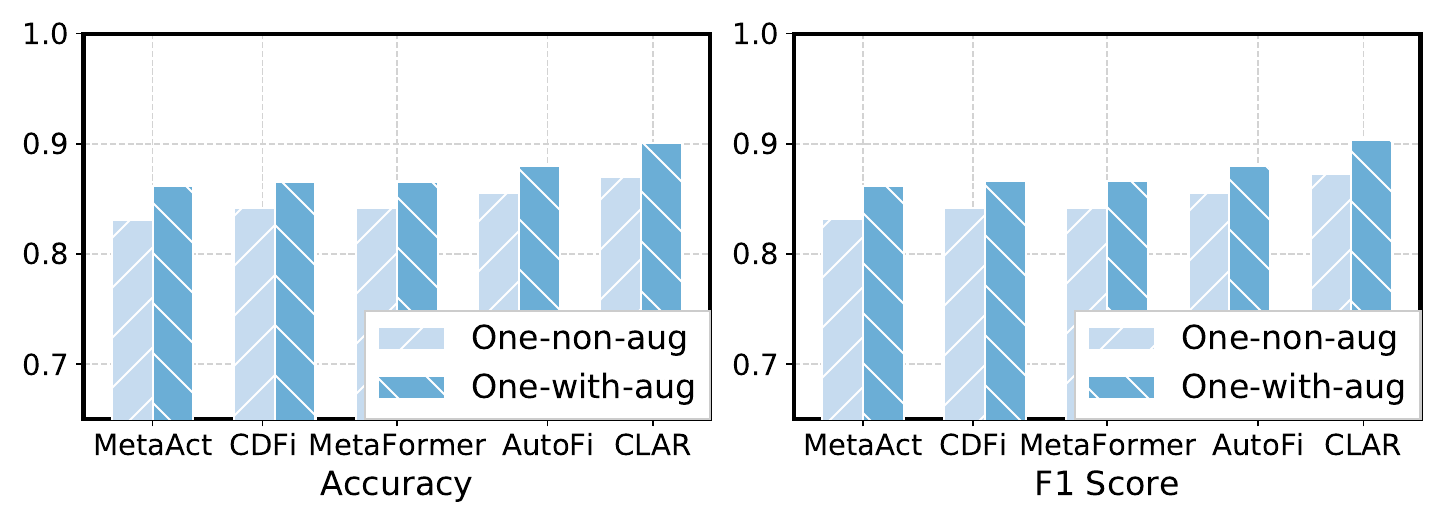}
	%	\vspace{-0.3cm}
	\caption{The performance with/without the augmented data for DeepSeg data.}
	\label{fig:DeepSegAug}
	\vspace{-0.2cm}
\end{figure}

Figure~\ref{fig:SignFiAug} and Figure~\ref{fig:DeepSegAug} show the accuracy and F1 for these four baselines and our CLAR with/without the augmented data generated by our DDPM-based augmentation model. As shown in Figure~\ref{fig:SignFiAug}, the performance of one-with-aug substantially exceeds that of one-non-aug for all the models on SignFi data. For example, the F1 of one-with-aug for AutoFi is about 3.6\% higher than that of one-non-aug.
%------------------------------rpt-------------------
%Figure~\ref{fig:SignFiAug} and Figure~\ref{fig:DeepSegAug} show the accuracy and F1 for these four baselines and our CLAR with/without augmented data. As shown in Figure~\ref{fig:SignFiAug}, one-with-aug substantially exceeds one-non-aug for these models on SignFi data. For example, the F1 of one-with-aug for AutoFi is about 3.6\% greater than one-non-aug.
%----
The DeepSeg dataset, presented in Figure~\ref{fig:DeepSegAug}, also exhibits similar trends. 
%These results indicate that our DDPM-based augmentation model can generate effective augmented samples by combining multiple samples, and further improve recognition performance. 
%------------------------------rpt-------------------
These results indicate our DDPM-based augmentation model can generate effective augmented samples by combining multiple samples, and advance identification performance. 
Also, our augmentation model can be applied to other similar recognition models.
Moreover, compared with the results in Figure~\ref{fig:SignFi}, there is a large performance degradation for our model CLAR. The reason behind is that the experiments in Figure~\ref{fig:SignFi} are under the scenario that the data of all the users is available for model training. While, in Figure~\ref{fig:SignFiAug}, the data of the test user is unseen during model training, making the learned model a little poor on the test user. 
However, our model significantly outperforms all the baselines, indicating that our model has higher generalization capacity compared to these baselines.

\subsection{Role of Labeled Data Size}
\label{subsec:label}

Our model requires a number of labeled data to fine-tune the classifier. Here we investigate the role of labeled data size. For these experiments, we select $p$ = [40, 60, 100]\% of the training samples as unlabeled data, and select $q$\% as the labeled data.
%------------------------------rpt-------------------
%To fine-tune the classifier, our method requires a certain number of labeled data. Here we conduct an investigation into the role of labeled data size. In these experiments, we select $p$ = [40, 60, 80]\% of the examples as unannotated data and $q$\% as the annotated data.

\begin{figure}[tb]
	\centering
	\includegraphics[width=0.48\textwidth]{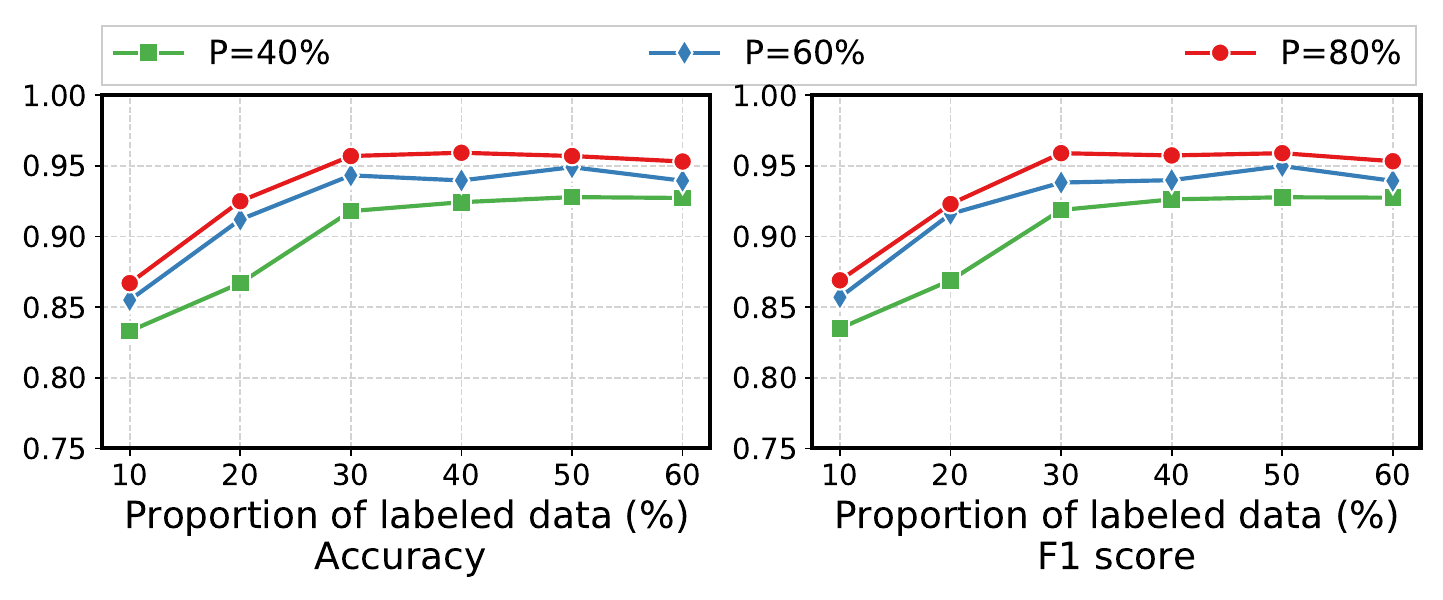}
	%\vspace{-0.3cm}
	\caption{The performance with different size of labeled data on SignFi data.}
	\label{fig:SignFidata}
	\vspace{-0.2cm}
\end{figure}

\begin{figure}[tb]
	\centering
	\includegraphics[width=0.48\textwidth]{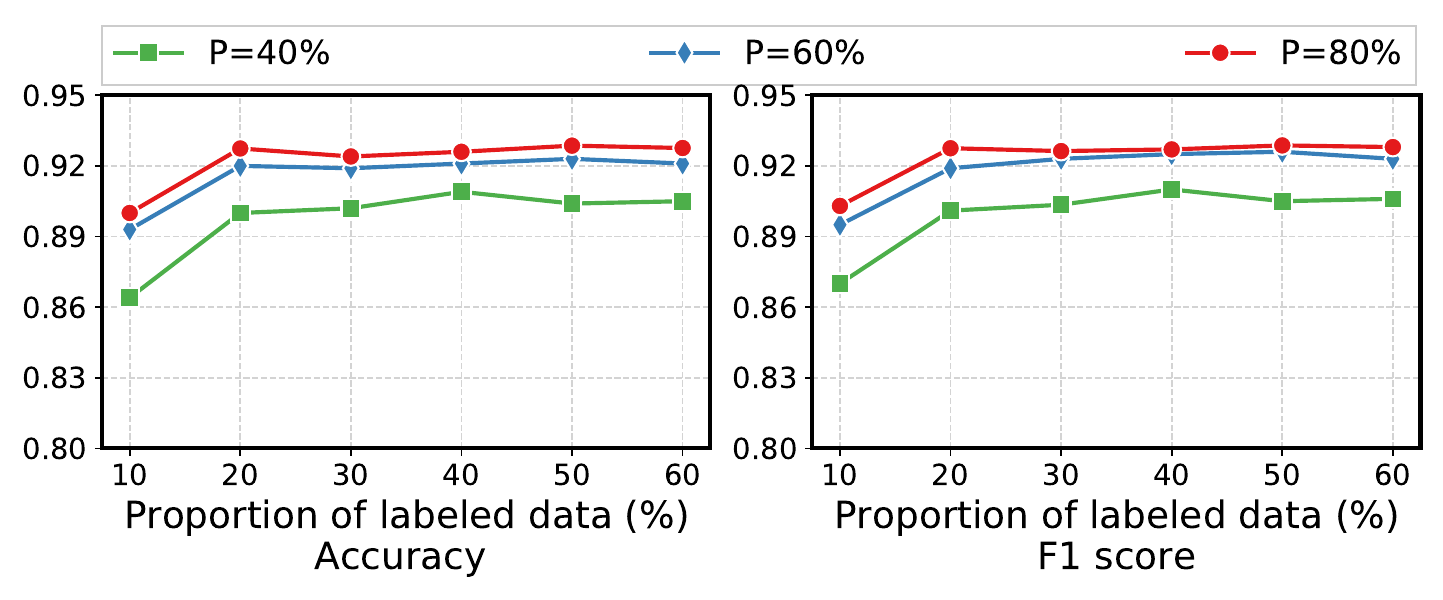}
	%	\vspace{-0.3cm}
	\caption{The performance with different size of labeled data on DeepSeg data.}
	\label{fig:DeepSegdata}
	\vspace{-0.2cm}
\end{figure}

As shown in Figure~\ref{fig:SignFidata} and Figure~\ref{fig:DeepSegdata}, for all the $p$ values, our model achieves increasing accuracy and F1 score with the rise of the labeled data size on both datasets, which indicates that the labeled data size has an important impact on recognition performance for our model.
%------------------------------rpt-------------------
%As illustrated in Figure~\ref{fig:SignFidata} and Figure~\ref{fig:DeepSegdata}, our model exhibits increasing accuracy and F1 score on both datasets as the labeled data size increases, for all $p$ values. These results suggest that the labeled data size has a crucial influence on the recognition performance of our method.
While, when selecting 30\% and 20\% of the labeled data on SignFi and DeepSeg, individually, the performance becomes stable on both datasets, i.e., their accuracies are almost the same with that at 60\%. This suggests that our model can efficiently take advantage of a few labeled samples to obtain the expected performance.

Moreover, the growth rates of the two datasets are different. The accuracies on SignFi data increase sharply, while they are relatively stable on DeepSeg data. The reason behind is because the number of labeled samples per class is very different for the two datasets. In fact, there are 10 and 30 labeled samples per category for SignFi data and DeepSeg data, individually. Hence, the same ratio means the various number of labeled data for these two datasets, which further leads to different performances. These results suggest that our model needs a certain amount of training data to achieve better performance. However, the number of labeled data, such as 10 per category, is easily affordable by human labeling.
%------------------------------rpt-------------------
%Furthermore, it is worth noting that the growth rates of the two datasets differ. The accuracies on SignFi experience a steep increase, while they remain comparatively stable on DeepSeg. This discrepancy can be attributed to the fact that the number of labeled samples per class varies greatly between the two datasets. Specifically, there are 10 and 30 labeled samples per category for SignFi and DeepSeg, respectively. Therefore, the same ratio results in different amounts of labeled data, which consequently affects their performances differently. These results emphasize the importance of providing our model with a few labeled samples. However, this number of labeled data can be easily obtained via human labeling.

\section{Related Work}
\label{sec:relate}

This work is mainly related to two research areas: CSI-based activity recognition and contrastive self-supervised learning. Here, we will present an overview of the most closely related works in each area.
%------------------------------rpt-------------------
%The study primarily is relevant to two domains: CSI-based activity recognition and contrastive self-supervised learning. In this section, we will provide an overview of the most relevant prior research.

%and emphasize the major distinctions between our research and theirs.

\subsection{CSI-based Activity Recognition}
\label{subsec:activity recognition}

The studies on WiFi CSI-based activity recognition can be divided into three genres according to the number of available labeled data: supervised and semi-supervised approaches and few-shot learning-based methods. 
%------------------------------rpt-------------------
%The research related to activity recognition using WiFi CSI can be categorized into three groups based on the availability of labeled data: supervised approaches, semi-supervised ways, and few-shot learning-based methods.

\textit{Supervised methods} mainly aim to adopt a number of labeled data to train classification models for activity identification. 
The researchers principally exploit different features and/or techniques to enhance recognition performance~\cite{yousefi2017survey,zhang2020device}.
For example, 
Chen \emph{et al.}~\cite{chen2021rf} exploit information of both time and frequency domains to build an end-to-end method for action identification, and introduce point-wise and depth-wise convolutions to boost the efficacy.
%apply point-wise grouped convolution and depth-wise separable convolutions to confine the model scale and speed up the inference execution time.
Zhang \emph{et al.}~\cite{zhang2021data} present a data augmentation method to transform and generate CSI data for alleviating the impact of action inconsistency and subject-specific issues and adopt a Dense-LSTM to classify activities.
%This paper proposes a multi-stage deep learning model consisting of a convolutional neural network and other popular deep neural architectures, namely Alexnet, Googlenet, and Squeezenet, for Wi-Fi sensing-based human activity recognition.
Sruthi \emph{et al.}~\cite{2022wifidl} illustrate a multi-stage deep learning framework, which is composed of multiple neural architectures, such as Alexnet, Googlenet and Squeezenet, for WiFi signal-based action recognition.
Xiao \emph{et al.}~\cite{DeepSeg2021} explore the interaction between the activity segmentation and classification to improve activity recognition performance.
%design an attention-based bi-directional long short-term memory model for passive human activity recognition using WiFi CSI signals.
Chen \emph{et al.}~\cite{chen2019wifi} design an attention-based Bi-LSTM model for WiFi CSI-based activity recognition.
%Shi \emph{et al.}~\cite{2022AFEE-MatNet} propose an innovative scheme, which combines an activity-related feature extraction and enhancement method and matching network. The proposed scheme can be directly applied in new/unseen environments without retraining.
Shi \emph{et al.}~\cite{2022AFEE-MatNet} extract action-related features to mitigate environment noises and improve the recognition performance under unseen scenes.

%Semi-supervised learning try to leverage unlabeled data to compensate for the shortage of labeled data for activity recognition~\cite{Bruno2018Semi}.
\textit{Semi-Supervised approaches} 
seek to use labeled as well as unlabeled data to perform classification tasks. 
It can leverage unlabeled data to compensate for the shortage of labeled data for activity recognition~\cite{Bruno2018Semi}.
For instance, 
Xiao \emph{et al.}~\cite{2019CsiGAN} propose a semi-supervised generative adversarial network to exploit unlabeled data for CSI-based activity recognition.
%Yuan \emph{et al.}~\cite{2023SSHAR} propose a human continuity activity semi-supervised recognizing method in multi-view IoT network scenarios. They combine supervised activity feature extraction with unsupervised encoder-decoder modules, which can capture continuity activity features from sensor data streams.
Wang \emph{et al.}~\cite{2021MCBAR} devise a multi-modal CSI-based activity recognition framework, which leverages a multimodal generator to amplify limited training data.
Yuan \emph{et al.}~\cite{2023SSHAR} design a semi-supervised action recognition model for multi-view Internet of Things scenes, which adopts an  encoder-decoder model to extract action features from sensing data for action recognition.
Recently, Mean Teacher has been introduced to build a semi-supervised activity recognition model using WiFi CSI~\cite{xiao2023mean}.

%SimCLR-based methods achieve state-of-the-art results in the fields of image processing.

\textit{Few-shot learning-based methods} intend to recognize a set of target classes by learning with sufficient labeled samples from a set of source classes but only with a few labeled samples from the target classes~\cite{wang2020generalizing}. Due to the difficulty of collecting numerous labeled data in the target domain, this technique is widely applied to the field of activity recognition.
%------------------------------rpt-------------------
%\textit{Few-shot learning-based methods} intend to identify a group of target categories by learning with enough annotated examples from a group of source categories, but with only a few annotated samples from the target ~\cite{wang2020generalizing}. 
This method is widely used in activity recognition due to the difficulty of gathering numerous annotated data in the target domain.
For instance, 
Wang \emph{et al.}~\cite{2021AFSLHAR} design a few shot learning-based action recognition approach, which can acquire the expected performance of identifying new types of activities using a few samples.
%to fine-tune the model parameters and avoid retraining the network from scratch.
Zhang \emph{et al.}~\cite{Across2022} and Ding \emph{et al.}~\cite{2021RF-Net} introduce meta-learning to conduct activity recognition, which can apply to new scenes or unseen actions by fine-tuning the model with very little training effort.
Zhang \emph{et al.}~\cite{2022CSI-GDAM} propose a graph-based few-shot learning framework with dual attention mechanisms for human activity recognition. The method adopts a convolutional block attention model to obtain activity related features from WiFi CSI. 
Shi \emph{et al.}~\cite{2022ERDF} design a human activity recognition scheme using the matching network with enhanced CSI to perform one-short learning for recognizing human activities in a new environment.
%------------------------------rpt-------------------
%Shi \emph{et al.}~\cite{2022ERDF} design a human activity recognition system that utilizes a matching network with refined CSI to achieve one-shot learning for recognizing actions in a novel scene.
Feng \emph{et al.}~\cite{2019Fslbar} propose a few-shot activity recognition method, which exploits a deep learning module for extracting features and implements knowledge transfer by transferring model parameters.

\subsection{Contrastive Self-Supervised Learning}
\label{subsec:contrastive}

%Contrastive Learning. Contrastive learning [27, 21, 41, 40] has become one of the most successful methods in the feld of self-supervised learning. As we mentioned, most recent works mainly focus on the augmentation for positive samples and the exploration for negative samples.
%----Weakly Supervised Contrastive Learning, iccv 2021

%Contrastive learning [27, 21, 41, 40] has become one of the most successful methods to learn item representations without requiring large-scale labeled data. A great amount of research has been initiated to enhance learned representations by designing different pretext tasks.

%----Anomaly Detection on Attributed Networks via Contrastive Self-Supervised Learning, tnnls 2021

%Through hand-crafted contrastive pretext tasks, these approaches learn representations by contrasting positive instance pairs against negative instance pairs~\cite{SimCLR2020}.

%----Dual-Stream Contrastive Learning for Channel State Information Based Human Activity Recognition
%A popular paradigm of contrastive learning methods is using different data augmentation to learn the invariant information underneath noisy data, which is typically achieved by maximizing the similarity of representations obtained from different data augmentations of a sample using the same network.

Contrastive learning is a significant division of self-supervised learning, which indicates the learning paradigm where supervision signals are generated from data itself~\cite{GenerorCon2021}.
%------------------------------rpt-------------------
%Contrastive learning is a significant division of self-supervised learning, a learning paradigm that derives its supervision signals from data itself~\cite{GenerorCon2021}.
It is a subset of unsupervised learning since it does not require any manual labels for model training.
%This technique tries to transform one item into multiple views, minimizes the distance between views from the same item, and maximizes the distance between views from different items in a feature map~\cite{SimCLR2020}.  
Contrastive methods have been applied to multiple fields, such as image processing, voice and natural language processing, and activity recognition.
%------------------------------rpt-------------------
%Contrastive methods have been applied to multiple domains, including image, voice and natural language processing, as well as activity recognition.

In the field of \textit{image processing}, various methods have been initiated to augment data and build effective views.
For example, SimCLR~\cite{SimCLR2020} proposes the composition of multiple data augmentations, e.g., Grayscale, Random Resized Cropping, Color Jittering, and Gaussian Blur, to make the model more robust.
%------------------------------rpt-------------------
%For example, SimCLR~\cite{SimCLR2020} proposes the composition of multiple data augmentations, e.g., Random Cropping, Color Jittering, and Gaussian Blur, to make the model more robust.
InfoMin~\cite{2019Contrastive} designs an information maximization principle that suggests a useful augmentation strategy should minimize the mutual information between positive pairs while preserving task-specific information. To explore the use of negative samples, InstDisc~\cite{Unsupervised2018} presents a memory bank to store the representation of all the images in the dataset.
%------------------------------rpt-------------------
%InfoMin~\cite{2019Contrastive} designs an information maximization principle that indicates a useful augmentation method should minimize the mutual information of positive pairs while preserving task-specific information. To take advantage of negative samples, InstDisc~\cite{Unsupervised2018} presents a memory bank to save the representations of all the examples into a data set.
%-----
%Meanwhile, SwAV~\cite{2020SwAV} proposes to compute cluster assignments online while enforcing consistency between cluster assignments obtained from views of the same image.
Meanwhile, SwAV~\cite{2020SwAV} introduces an online cluster assignment calculation approach while simultaneously ensuring consistency between the cluster assignments obtained from multiple views of the same image.
%MoCo~\cite{2020MoCo} increases the number of negatives by using a momentum contrast mechanism that forces the query encoder to learn the representation from a slowly progressing key encoder and maintains a long queue to provide a large number of negative examples.
MoCo~\cite{2020MoCo} proposes an approach to enhance the number of negative examples by utilizing a momentum contrast mechanism. This method compels the query encoder to learn the representation from a slowly updating key encoder, while maintaining a long queue of samples to offer a substantial number of negative examples.
%SupCon~\cite{2020Supervised} shows that the positive and negative instances created by SimCLR do not take into account the correlation of features between different pictures belonging to the same class. Clusters of points belonging to the same class are pulled together in embedding space, while simultaneously pushing apart clusters of samples from different classes.
SupCon~\cite{2020Supervised} shows that the positive and negative instances created by SimCLR do not take into account the correlation of features between different pictures belonging to the same class. In contrast, SupCon pulls together clusters of points from the same category in the embedding space while pushing apart clusters of samples from different categories.
%WCL~\cite{2021WCL} proposes a k-nearest neighbor based multi-crops strategy. They store the feature for every batch and then use these features to find the $K$ closest samples based on the cosine similarity at the end of each epoch.
WCL~\cite{2021WCL} introduces a multi-crops strategy based on $k$-nearest neighbors. It saves the features for each batch and utilizes them to determine the $k$ nearest samples using cosine similarity at the end of every epoch.
CLSA~\cite{2022CLSA} proposes to build a stronger augmentation by a random combination of different augmentations.

In addition, contrastive learning is also widely used for \textit{voice and natural language processing}.
For example, in the field of voice processing,
Yakura \emph{et al.}~\cite{2022SSCLSV} introduce self-supervised contrastive learning to obtain feature representations of singing voices. 
%------------------------------rpt-------------------
%Yakura \emph{et al.}~\cite{2022SSCLSV} introduce contrastive learning to obtain the voice representation. 
Tang \emph{et al.}~\cite{2022AVQVC} propose a new one-shot voice conversion model that utilizes vector quantization voice conversion to separate the content and speaker information.
%In the domain of natural language processing,
For natural language processing,
%Qin \emph{et al.}~\cite{2022GL-CLeF} utilize contrastive learning to explicitly align similar representations across source language and target language.
Qin \emph{et al.}~\cite{2022GL-CLeF} employ contrastive learning to align similar representations between the source and target languages explicitly.
%Han \emph{et al.}~\cite{2022CLFGET} propose a cross-lingual contrastive learning framework to learn FGET models for low-resource languages.
Han \emph{et al.}~\cite{2022CLFGET} introduce a cross-lingual contrastive learning method for learning FGET models from low-resource languages.

Recently, contrastive learning is adopted to enhance the performance of \textit{sensor-based activity recognition}.
For example, Jain \emph{et al.}~\cite{2022ColloSSL} introduce a collaborative self-supervised learning method for activity recognition using sensors, which utilizes natural transformations in the sensor datasets gathered from multiple devices to conduct contrastive learning.
%------------------------------rpt-------------------
%For example, Jain \emph{et al.}~\cite{2022ColloSSL} introduce a collaborative contrastive learning method for action identification using sensor data.
%Khaertdinov \emph{et al.}~\cite{khaertdinov2021contrastive} combine a transformer-based encoder into a contrastive self-supervised learning framework to learn effective feature representations for sensor-based human activity recognition.
Khaertdinov \emph{et al.}~\cite{khaertdinov2021contrastive} integrate a transformer-based encoder into a contrastive self-supervised learning model to learn the efficient feature representation for activity recognition using sensor data.
%Haresamudram \emph{et al.}~\cite{haresamudram2020masked} introduce masked reconstruction as a viable self-supervised pre-training objective for wearable sensing device-based human activity recognition.
Haresamudram \emph{et al.}~\cite{haresamudram2020masked} propose masked reconstruction as a suitable self-supervised pre-training objective for  action recognition using wearable sensing devices.
%Saeed \emph{et al.}~\cite{2019MulSSL} design a multi-task self-supervised approach, which presents a multi-task temporal convolutional network to learn generalizable features from sensory data.
Saeed \emph{et al.}~\cite{2019MulSSL} propose a multi-task self-supervised approach that employs a multi-task temporal convolutional network to learn generalizable information from sensing data.
%Xu \emph{et al.}~\cite{xu2023dual} design a dual-stream contrastive learning model that can process and learn the raw WiFi CSI data in a self-supervised manner. 
Xu \emph{et al.}~\cite{xu2023dual} design a dual-stream contrastive learning method which processes and learns the raw WiFi CSI data by a self-supervised way. 
%Liu \emph{et al.}~\cite{liu2021contrastive} introduce a short-time fourier neural network-based contrastive self-supervised representation learning framework, which takes both time-domain and frequency-domain features into consideration. 
Liu \emph{et al.}~\cite{liu2021contrastive} present a framework for contrastive self-supervised representation learning based on a short-time Fourier neural network, which incorporates both time-domain and frequency-domain features.
%Koo \emph{et al.}~\cite{koo2022contrastive} devise a self-supervised learning task that pairs the accelerometer and the gyroscope embeddings acquired from the same activity instance. 
Koo \emph{et al.}~\cite{koo2022contrastive} devise a self-supervised learning framework, which involves pairing accelerometer and gyroscope embeddings obtained from the same activity sample for model training.
Yang \emph{et al.}~\cite{yang2023autofi} illustrate a self-supervised learning activity recognition model using WiFi CSI, which can transfer the knowledge learned from unlabeled low-quality CSI samples to specific domains.
%------------------------------rpt-------------------
%Yang \emph{et al.}~\cite{yang2023autofi} illustrate a self-supervised learning action identification model utilizing CSI, which can transfer the knowledge learned from unlabeled low-quality CSI samples to specific domains.
%Wang \emph{et al.}~\cite{wang2022sensor} present a sensor data augmentation approach for contrastive learning, which considers variable domain information and simulates realistic activity data by changing the sampling frequency to maximize the coverage of the sampling space. 
Wang \emph{et al.}~\cite{wang2022sensor} present a sensor data augmentation approach for contrastive learning, which generates augmented data by changing the sampling frequency. The augmented data can enlarge the sampling space to improve recognition performance.
Wang \emph{et al.}~\cite{wang2023negative} present a new contrastive learning framework for sensor-based human activity recognition, which first clusters the instance representations, and then regards samples from different clusters as negative pairs.

\section{Conclusions}
\label{sec:conclusion}

In this paper, we presented a diffusion model-based Contrastive Learning framework for human Activity Recognition (CLAR) using WiFi CSI. In this framework, we designed a DDPM-based time series-specific augmentation model, which imposes the high-frequency and low-frequency features from the reference sample with varying weights on the reverse diffusion process to generate effective augmented samples. Also, we presented an adaptive weight algorithm, which adaptively adjusts the weights of positive sample pairs for learning better data representations. Based on the two public datasets, the experimental results illustrate that CLAR significantly outperforms the state-of-the-art baselines. 

\bibliographystyle{IEEEtran}
\bibliography{IEEEexample1}

%This work is supported by the National Natural Science Foundation of China (No.62072077), Science and Technology Foundation of Henan Province of China (No.212102210387). 

\begin{IEEEbiography}[{\includegraphics[width=1in,height=1.25in,clip,keepaspectratio]{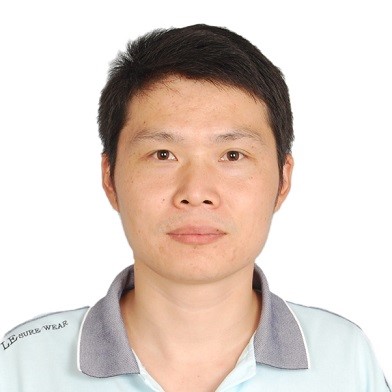}}]{Chunjing Xiao}
received the Ph.D. degree in computer software and theory from the University of Electronic Science and Technology of China, Chengdu, Sichuan, China, in 2013.

He is currently an Associate Professor with the School of Computer and Information Engineering, Henan University, Kaifeng, China. He was a Visiting Scholar with the Department of Electrical Engineering and Computer Science, Northwestern University, Evanston, IL, USA. His current research interests include Internet of Things, anomaly detection, and recommender systems.
\end{IEEEbiography}

\begin{IEEEbiography}[{\includegraphics[width=1in,height=1.25in,clip,keepaspectratio]{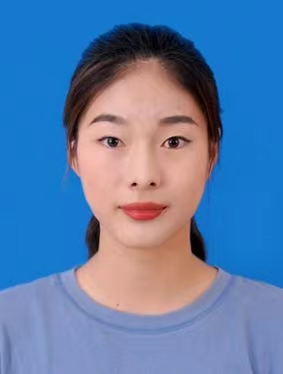}}]{Yanhui Han}
received the M.S degree in  Electronic Information from the School of Computer and Information Engineering at Henan University in Kaifeng, China, in 2024.

Her research interests include Internet of Things, wireless networks and data analytics.
\end{IEEEbiography}

\begin{IEEEbiography}[{\includegraphics[width=1in,height=1.25in,clip,keepaspectratio]{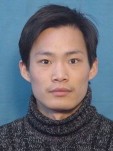}}]{Wei Yang}
received Ph.D. degree in Computer Application Technology from Harbin Institute of Technology in 2011. 

He is currently an Associate professor in School of Computer and Information Engineering, Henan University, Kaifeng, China. His research interests include machine learning, deep learning, metric learning and bioinformatic.
\end{IEEEbiography}

\begin{IEEEbiography}[{\includegraphics[width=1in,height=1.25in,clip,keepaspectratio]{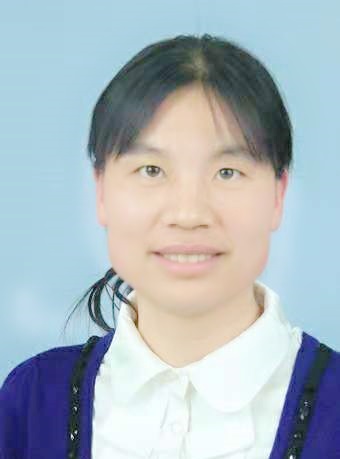}}]{Yan-e Hou}
received the Ph.D. degree in Remote Sensing Information Science and Technology
from Henan University, Kaifeng, China, in 2016.

She is currently an Associate Professor with the School of Computer and Information Engineering, Henan University, Kaifeng, China. Her research interests currently include intelligent optimization algorithms, artificial intelligent and its relative applications.
\end{IEEEbiography}

\begin{IEEEbiography}[{\includegraphics[width=1in,height=1.25in,clip,keepaspectratio]{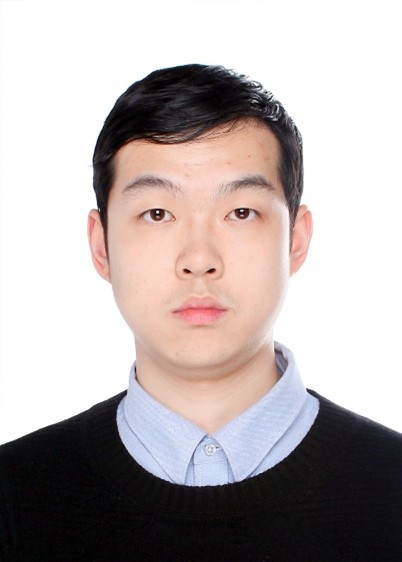}}]{Fangzhan Shi}
received the B.Eng. in Telecommunication Engineering in 2017 and M.Sc. in Robotics in 2018 at Hangzhou Dianzi University, China and University College London respectively. 

He worked as an artificial intelligence engineer at Supcon, China in 2019 and 2020.  He is currently a PhD student in the department of security and crime science, University College London. His research interest is joint communication and sensing.
\end{IEEEbiography}

\begin{IEEEbiography}[{\includegraphics[width=1in,height=1.25in,clip,keepaspectratio]{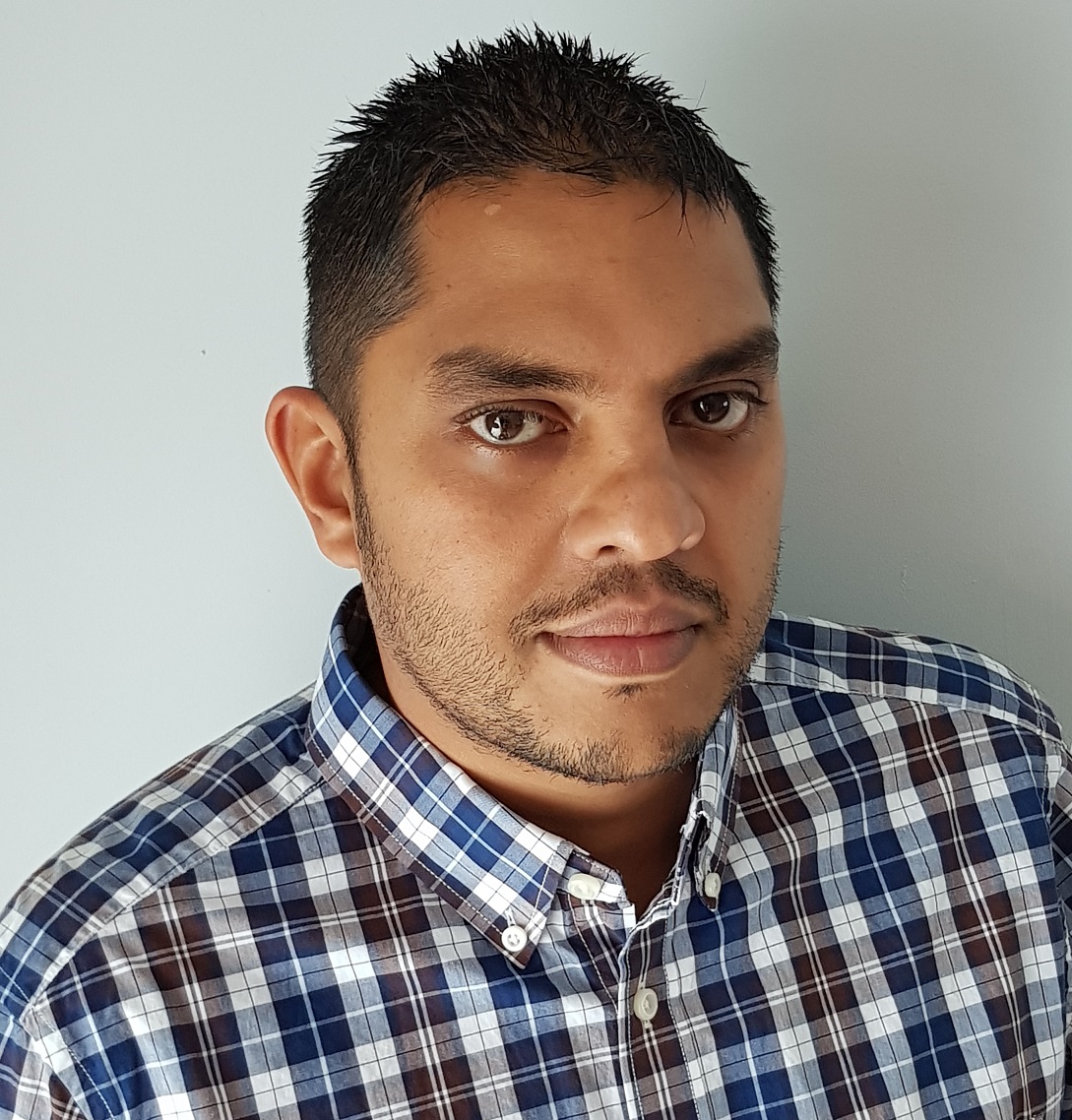}}]{Kevin Chetty}
received the B.Sc. degree in physics degree from King’s College London in 2003, the MRes degree in  physics, and the Ph.D. degree in medical ultrasound imaging from Imperial College London in 2007.

He is an Associate Professor at University College London where he leads the Urban Wireless Sensing Lab. He has pioneered work in passive WiFi sensing; an area of sensing research expected to drive advancements in ubiquitous sensing and smart environments. Dr. Chetty has developed patented techniques for high-throughput data processing in passive wireless systems to facilitate real-time operation, and demonstrated the first through-the-wall detections using the technology. He has over 100 conference and journal publications in the application of sensing systems and signal processing techniques for situational awareness and human behaviour classification using micro-Doppler signatures, machine learning and software-defined sensors. 
\end{IEEEbiography}

\end{document}